\documentclass[letterpaper]{emulateapj}
\usepackage{apjfonts}

\newcommand{\hl}{} 
\newcommand{\fig}[1]{\mbox{Fig.~\ref{#1}}}
\newcommand{\sect}[1]{\mbox{\S~\ref{#1}}}
\newcommand{\tab}[1]{\mbox{Table~\ref{#1}}}

\newcommand{\ang}{\mbox{\AA}}
\newcommand{\km}{\mbox{km}}

\newcommand{\s}{\mbox{s}}
\newcommand{\yr}{\mbox{yr}}
\newcommand{\Myr}{\mbox{Myr}}
\newcommand{\Gyr}{\mbox{Gyr}}

\newcommand{\kpc}{\mbox{kpc}}
\newcommand{\Mpc}{\mbox{Mpc}}

\newcommand{\Jy}{\mbox{Jy}}

\newcommand{\Ha}{\mbox{H$\alpha$}}
\newcommand{\Hb}{\mbox{H$\beta$}}
\newcommand{\Hc}{\mbox{H$\gamma$}}
\newcommand{\Hd}{\mbox{H$\delta$}}

\newcommand{\HII}{\mbox{H\,{\sc ii}}}
\newcommand{\HeI}{\mbox{He\,{\sc i}}}
\newcommand{\Mg}{\mbox{Mg}}
\newcommand{\MgIb}{\mbox{Mg\,{\sc i}~b}}
\newcommand{\MgII}{\mbox{Mg\,{\sc ii}}}

\newcommand{\fNII}{\mbox{[N\,{\sc ii}]}}
\newcommand{\fOIII}{\mbox{[O\,{\sc iii}]}}
\newcommand{\Na}{\mbox{Na}}
\newcommand{\NaI}{\mbox{Na\,{\sc i}}}

\newcommand{\NaID}{\mbox{Na\,{\sc i}~D}}

\newcommand{\eV}{\mbox{eV}}
\newcommand{\lam}{\mbox{$\lambda$}}

\newcommand{\chisq}{\mbox{$\chi^2$}}
\newcommand{\snr}{\mbox{S/N}}
\newcommand{\snrpp}{\mbox{$(\snr)_{\rm pix}$}}

\newcommand{\LCDM}{\mbox{$\Lambda$CDM}}

\newcommand{\Cf}{\mbox{$C_f$}}

\newcommand{\NUV}{\mbox{NUV}}

\newcommand{\Lir}{\mbox{$L_{\rm IR}$}}
\newcommand{\Msun}{\mbox{$M_\odot$}}

\newcommand{\pblue}{\mbox{$p_{\rm blue}$}}

\newcommand{\scalefigure}{} 

\shorttitle{LOW-IONIZATION OUTFLOWS IN AEGIS}
\shortauthors{SATO ET AL.}

\begin{document}

\title{AEGIS: The Nature of the Host Galaxies of Low-ionization
Outflows at $z < 0.6$}

\author{Taro Sato\altaffilmark{1,2},
Crystal L. Martin\altaffilmark{2,3},
Kai G. Noeske\altaffilmark{4},
David C. Koo\altaffilmark{5},
Jennifer M. Lotz\altaffilmark{6}
}

\altaffiltext{1}{nomo17k@gmail.com}

\altaffiltext{2}{Department of Physics, University of California,
Santa Barbara, CA 93106-9530, U.S.A.}

\altaffiltext{3}{Packard Fellow}

\altaffiltext{4}{Keck Foundation Fellow; Harvard-Smithsonian Center
for Astrophysics, Cambridge, MA, U.S.A.}

\altaffiltext{5}{UCO/Lick Observatory, Department of Astronomy \&
Astrophysics, University of California, Santa Cruz, U.S.A.}

\altaffiltext{6}{Leo Goldberg Fellow; National Optical Astronomical
Observatories, Tucson, AZ, U.S.A}

\begin{abstract}
We report on a signal-to-noise (\snr) limited search for
low-ionization gas outflows in the spectra of the $0.11 < z < 0.54$
objects in the Extended Groth Strip (EGS) portion of the Deep
Extragalactic Evolutionary Probe 2 (DEEP2) survey.  Doppler shifts
from the host galaxy redshifts are systematically searched for in the
$\NaI~\lam~5890,96$ doublet (\NaID).  Although the spectral resolution
and \snr\ limit us to study the interstellar gas kinematics from
fitting a single doublet component to each observed \NaID\ profile,
the typical outflow often seen in local luminous-infrared galaxies
(LIRGs) should be detected at $\ga 6\sigma$ in absorption equivalent
width down to the survey limiting \snr\ ($\sim 5$ pixel$^{-1}$) in the
continuum around \NaID.  The detection rate of LIRG-like outflow
clearly shows an increasing trend with star-forming activity and
infrared luminosity.  However, by virtue of not selecting our sample
on star formation, we also find a majority of outflows in galaxies on
the red sequence in the rest-frame ($U-B$, $M_B$) color-magnitude
diagram.  Most of these red-sequence galaxies hosting outflows are of
early-type morphology and show the sign of recent star formation in
their UV-optical colors; some show enhanced Balmer \Hb\ absorption
lines indicative of poststarburst as well as high dust extinction.
These findings demonstrate that outflows outlive starbursts and
suggest that galactic-scale outflows play a role in quenching star
formation in the host galaxies on their way to the red sequence.  The
fate of relic winds, as well as the observational constraints on
gaseous feedback models, may be studied in galaxies during their
poststarburst phase.  We also note the presence of inflow candidates
in red, early-type galaxies, some with signs of active galactic
nuclei/LINERs but little evidence for star formation.
\end{abstract}


\journalinfo{To appear in Astrophysical Journal}
\received{April 27, 2008}
\accepted{February 9, 2009}

\section{Introduction}

A number of large-scale galaxy surveys of unprecedented volume and
depth are steadily mapping out the luminous constituents of the
universe out to higher redshifts.  The lambda cold dark matter (\LCDM)
cosmological paradigm describes the gravity-driven assembly history of
the universe quite successfully on the largest physical scales.  Yet
our understanding of the universe remains largely incomplete on the
galaxy scales where the astrophysical processes involving stellar
evolution and feedback of interstellar and intergalactic matter are
essential --- baryons can cycle through these in an intricate,
multiphase manner.  Gaseous feedback processes have been gaining
serious attention, since theoretical predictions of the growth of
luminous structure are highly sensitive to the ``prescriptions'' for
these small-scale, subgrid physics \citep[e.g.,][]{crot06, bowe06,
hopk06}, yet the observational constraints are still relatively
scarce.

Absorption-line probes of the gas entrained in galactic ``superwinds''
\citep[e.g.,][]{chev85, heck90} have been effective in constraining
how much matter could be carried by outflows through the lines in the
rest-frame optical \citep[e.g.,][]{heck00, rupk02, rupk05p2, mart05}
and ultraviolet \citep[e.g.,][]{heck01, schw06}, which may pollute
intergalactic media and eventually deplete host galaxies of fuel for
further star formation.  The low-ionization absorption line studies
suggest that a considerable amount of neutral gas may be carried away
by outflows, $\sim 10^4$--$10^7~\Msun$ in dwarfs \citep{schw04} and
$\sim 10^8$--$10^{10}~\Msun$ in ultraluminous infrared galaxies
\citep[ULIRGs;][]{mart05, rupk05p2}.  The column density of outflowing
cool matter, however, can only be estimated after making a series of
crude assumptions on its geometry, ionization state, dust-depletion
factor, and metallicity.  Although the reliability of such mass
outflow estimates could therefore be questioned, an unambiguous
detection of outflow is relatively straightforward in finding a
Doppler-shifted absorption component in a line complex of interest.
So far, a vast majority of existing studies of outflows have focused
on galaxies selected a priori to have high star formation rates
(SFRs), from dwarf starburst galaxies \citep[e.g.,][]{schw04} and
luminous infrared galaxies \citep[e.g.,][]{mart05, mart06, rupk05p1,
rupk05p2} in the local universe to Lyman-break galaxies at high
redshifts \citep[e.g.,][]{shap03}; see \citet{veil05} for a recent
review.  Although winds have been unambiguously detected in these
systems, much work remains to be done in characterizing outflows in
terms of their host galaxy properties in a large, relatively unbiased
sample.

The $\NaI~\lam~5890,96$ doublet (\NaID) absorption can be stellar or
interstellar in origin yet is a useful line for a census of cool winds
in the interstellar medium (ISM).  Since it is an absorption line
measured against the background light of host galaxies, its Doppler
shift relative to the systemic redshift can be measured reasonably
well.  The \NaID\ absorption line directly traces cool ($T \sim 100$
K) gas which may directly fuel star formation.  In (U)LIRGs, $\sim
10\%$ of the dynamical mass can be entrained in cool outflows
\citep{rupk05p2, mart06}, suggesting that the bulk of outflowing mass
resides at the cool phase.  The \NaID\ doublet is a resonance line and
forms among the most prominent absorption lines detectable out to $z
\sim 0.5$ in optical spectra.

There is a pressing need to extend the study of outflows to the
systems with lower SFRs as well as at later stages of star formation.
For one thing, the observed spatial extent of outflows in local ULIRGs
and a simple dynamical argument suggest that outflowing gas clouds may
outlive starbursts or active galactic nuclei (AGNs) which may drive
outflows \citep{mart06}.  Recently, \citet{trem07} found that $z \sim
0.5$ massive poststarburst galaxies, observed at up to $1.5~\Gyr$
after intense episodes of star formation, host outflows almost as fast
as or even faster than those observed in local ULIRGs.  Since outflows
may carry away from galaxies a significant fraction of gas mass that
would otherwise be available for further star formation, knowing the
``fate'' of outflowing matter is clearly of interest.

In this paper, we present the result from a systematic search for
\NaID\ outflows in a flux-limited sample of galaxies, drawn from the
Extended Groth Strip (EGS) portion of the Deep Extragalactic
Evolutionary Probe 2 survey \citep[DEEP2;][]{davi03}.  Since the EGS
field has been extensively observed by a wide array of
multiwavelength missions as part of the All-wavelength Extended Groth
Strip Survey \citep[AEGIS;][]{davi07}, we are able to characterize our
spectroscopically selected sample in view of various physical
quantities.  Although \NaID\ is a resonance line and among the most
prominent of absorption lines in the visible, rigorous absorption
analysis needs high signal-to-noise (\snr) continuum, which is a
requirement not quite satisfied by the vast majority of 1 hr
integration spectra from the DEEP2 survey.  Co-adding a number of
low-\snr\ spectra from a set of subsamples to improve the effective
\snr\ is a sensible approach.  Yet even such stacking analyses are
limited a priori by numerous ways in which subsamples can be
constructed; some prior knowledge must be gained about the subsampling
schemes in which the desired information can best be elucidated.  In
order to initiate an effort to carry out an unbiased census of
outflows in modern spectroscopic galaxy surveys, as well as to
motivate ensuing stacking analysis with proper subsampling schemes, it
is still beneficial to take an approach to seek evidence of outflows
in individual spectra.

An ambitious goal would be to estimate the quantities that are useful
for improving the prescriptions of cosmological semianalytical
simulations, such as outflow detection rate and mass-loading factor in
a robust volume-limited sample.  In this paper, however, our focus is
on simply characterizing the property of galaxies that host LIRG-like
outflows, using a rich set of multiwavelength observations from the
AEGIS survey.  We first describe the data, selection, and analysis
method for the \NaID\ sample in \sect{Data and Analysis}.  The host
galaxies of outflows, as defined from the \NaID\ kinematics, are then
studied in view of their UV, optical, and infrared properties in
\sect{Host Galaxies of Outflows}.  We then note a few caveats, frame
our findings in the larger context of galaxy evolution
(\sect{Discussion}), and summarize the paper (\sect{Summary}).
Throughout this paper, we adopt the standard \LCDM\ cosmology,
$(\Omega_m, \Omega_\Lambda) = (0.3, 0.7)$, with $H_0 =
70~\km~\s^{-1}~\Mpc^{-1}$.

\section{Data and Analysis}
\label{Data and Analysis}

\subsection{DEEP2 Spectra}

The EGS, covering $\approx 0.5~{\rm deg}^{2}$, is one of the four
fields observed in the DEEP2 survey \citep[see][for the descriptions
of the survey]{coil04, davi03, davi04}, in which the spectroscopic
targets are preselected by the Canada-France-Hawaii Telescope (CFHT)
$BRI$ photometry \citep{cuil01}.  While the targets in the other three
fields are preselected to the apparent magnitude limit of $R_{\rm AB}
\le 24.1$ and by color cuts to the galaxies likely to be at $z \ga
0.7$, such a color cut was used in the EGS to slightly down-weight
galaxies at $z \la 0.7$, resulting by design in roughly equal numbers
of galaxies above and below $z = 0.7$.  The spectra for $\simeq
13,570$ EGS objects are obtained with the DEIMOS spectrograph on the
Keck II telescope \citep{fabe03}.  The $1200$ lines mm$^{-1}$ grating
centered at $7800~\ang$ and the OG550 order-blocking filter with $1''$
slit widths are used, which leads to the spectral coverage roughly of
$6500$--$9100~\ang$.  The spectral resolution in FWHM is $\sim
68~\km~\s^{-1}$.  The data are processed by an automated pipeline to
produce (unfluxed) one-dimensional spectra (Newman et al., in
preparation).  The DEEP2 DEIMOS
pipeline\footnote{http://astro.berkeley.edu/\~{}cooper/deep/spec2d/}
produces an inverse-variance vector for each spectrum.  In this paper,
we use the spectra extracted via a variant of the optimal extraction
algorithm presented in \citet{horn86}.  Throughout this paper, we only
use spectra which are tagged DEEP2 ZQUALITY flag of four.

\subsection{Systemic Redshifts}
\label{Systemic Redshifts}

An outflow velocity is measured relative to a systemic redshift of a
host galaxy.  A major contribution to the uncertainty of outflow
velocity therefore comes from the uncertainty in the systemic redshift
measurement.  Each DEEP2 spectrum is assigned a spectroscopic redshift
from \chisq\ minimization with a few template spectra and has the
accuracy well quantified from repeat observations
\citep[$30~\km~\s^{-1}$ in RMS;][]{will06}.  Each redshift measurement
has been visually inspected and assigned a redshift quality
flag.\footnote{See http://deep.berkeley.edu/ for the DEEP2 survey
detail.}  Nevertheless, we carry out independent redshift measurements
using IRAF software package XCSAO \citep{kurt98}.  The primary
motivation for an independent set of redshift measurements is to mask
the \NaID\ complex.  The \NaID\ absorption line is among the most
prominent features in the visible part of the galaxy spectrum.  In the
case where a systemic redshift should only reflect the centroid of the
stellar motions, \NaID\ should be excluded from cross correlation with
stellar spectral templates, in order to avoid the interstellar
components, which may be redshifted or blueshifted from the systemic
redshift of the galaxy, to affect the result.

We adapt the cross-correlation templates (ID: 24--30) from the fifth
data release of the Sloan Digital Sky Survey \citep[SDSS;][]{adel06}.
The template which yields the smallest uncertainty is generally picked
and designated as the systemic redshift of the galaxy.  An uncertainty
in each redshift measurement is derived from the $r$ statistics, which
roughly corresponds to the \snr\ of cross-correlation peak from which
the best redshift estimate is computed \citep{tonr79}.  The agreement
between XCSAO and DEEP2 redshifts is generally good; for most
purposes, the systematic redshift disparity of $cz({\rm XCSAO}) -
cz({\rm DEEP2})\approx 10~\km~\s^{-1}$ is much less than the
instrumental resolution and is insignificant.  Where precise systemic
redshifts are critical, we use XCSAO measurements, since the
uncertainty is empirically calibrated and well understood among our
sample.  The redshift measurement stored in each DEEP2 spectrum almost
certainly underestimates the uncertainty in case of gross template
mismatch, since only three components are used to generate a template
for cross correlation, and $\chi^2$ statistics was computed after
continua are removed from them.  Each redshift measurement for which
$\left| cz({\rm XCSAO}) - cz({\rm DEEP2}) \right| > 2\sigma_v({\rm
XCSAO})$ is visually inspected, and the spectrum is either reassigned
a redshift from another template or removed from the sample.  Only one
object dropped out of the sample because of the failure in the
cross-correlation redshift measurement.

\subsubsection{Spectral Line Indices}
\label{Spectral Line Indices}

\begin{figure*}
\scalefigure
\plotone{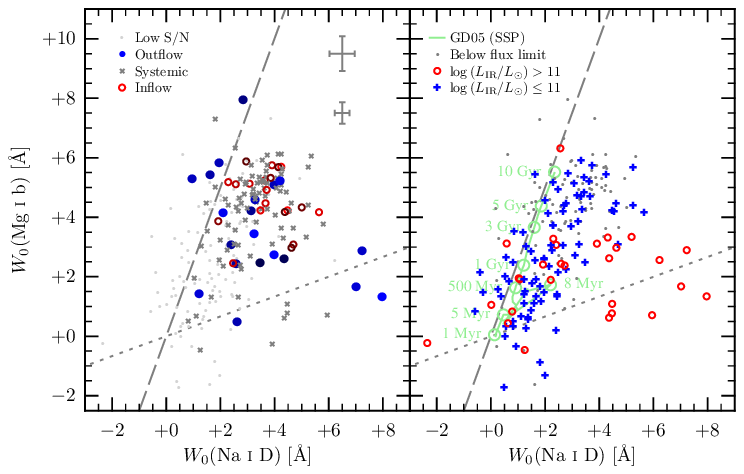}
\caption{ \MgIb-\NaID\ spectral index plane for the objects with
the measurements of \NaID\ velocity (\emph{left}) and the infrared
luminosity (\emph{right}).  See \sect{Spectral Line Indices} for the
definition of spectral indices.  In the left panel, the points are
denoted by \NaID\ kinematics into outflow (\emph{blue filled circle}),
systemic (\emph{gray cross}), inflow (\emph{red open circle}), and
low \snr\ (\emph{light gray dot}); see \sect{NaID Velocity and
Blueshift Probability} for the detail on \NaID\ kinematics.
Outflows/inflow data points are color-mapped with the blueshift
probability (\sect{NaID Velocity and Blueshift Probability}), where
bluer (redder) marks indicate stronger blueshift (redshift) of \NaID\
absorption line; see the inset of \fig{fig.dv_nad_vs_csnr_w_inset} for
the exact color-mapping scheme with the blueshift probability.  The
error bars shown for the spectral indices are the medians for high- and
low-\snr\ samples.  In the right panel, the symbols indicate LIRGs
(\emph{red open circle}), non-LIRGs (\emph{blue crosses}), and objects
below flux limit of the \emph{Spitzer}/MIPS $24~ \mu m$ survey
(\emph{gray dot}).  \citet{rupk05p1} found most of their
infrared-selected objects lying below the gray dotted line, $W(\NaID)
= 3W(\MgIb)$, to host outflows, whereas most objects above did not.
The long gray dashed line indicates the fiducial stellar loci,
$W(\NaID) = 0.4 W(\MgIb)$, where the slope was chosen to roughly match
the evolutionary track of the solar-metallicity, single stellar
population model from \citet[][\emph{solid green line}]{delg05}.  For
each object the differences between the observed and fiducial stellar
\NaID\ indices presumably indicate the interstellar contribution (see
\fig{fig.v_nad_vs_sfrac_nad_w_outflow_frac}).  (\emph{A color version of this figure is available in the online journal.})}
\label{fig.index_mgb_vs_index_nad}
\end{figure*}

Using the rest-frame band definitions in \tab{tab.line_index}, we
measure the spectral indices (denoted $W_0$, with the subscript
``$0$'' meaning rest frame) of some lines as estimates of their
equivalent widths.  Each spectral line index is computed via the
``flux-summing'' method.  First, a straight line ``pseudocontinuum''
is fitted to the variance-weighted pixel flux values from the blue and
red straddling continua as defined in \tab{tab.line_index}, via the
standard Levenberg-Marquardt algorithm.  The covariance matrix is used
to estimate the uncertainty in the pseudocontinuum.  Then at each
pixel $i$ with a pixel width $\Delta\lambda_i$ in wavelength falling
within the spectral bandpass defined as ``line'' in
\tab{tab.line_index}, the observed flux $f_o$ and the pseudocontinuum
flux $f_c$ is used to compute the flux excess or depletion, such that
\[
W = -\sum_i \left( \frac{f_o - f_c}{f_c} \right)_i \Delta\lambda_i
\]
yields the equivalent width index in the observed frame.  The
rest-frame value is computed simply from $W_0 = W / (1 + z)$ where $z$
is the redshift of the galaxy.  The uncertainty in the index is
estimated by formal propagation of uncertainties in $f_o$ and $f_c$
using the above relation.  By construction, $W_0 > 0~\ang$ ($<
0~\ang$) corresponds to a line flux seen in absorption (emission),
although the physical interpretation is slightly complicated by the
fact that both emission and absorption may exist in a line feature,
sometimes called emission filling.  Although not perfect, a line index
gives a good estimate of true equivalent width, when emission filling
is not significant.

Previous studies have used the equivalent width of the
$\MgIb\lam5167,73,85$ triplet to estimate the stellar contribution to
the \NaID\ absorption line \citep{heck00, rupk02, schw04, rupk05p1,
mart05}.  The correlation between \NaID\ and \MgIb\ equivalent widths
in stellar spectra is expected based on the similar mechanisms from
which \Na\ and \Mg\ are produced in stars and their roughly similar
ionization potentials ($5.14~\eV$ and $7.65~\eV$ for \Na\ and \Mg,
respectively).  \citet{heck00}, \citet{mart05}, and \citet{rupk05p1}
all showed high equivalent width ratios of \NaID\ to \MgIb\ to be a
good indicator of the presence of winds from their samples of
infrared-luminous galaxies.  This assumption is reasonable provided
that a presence of a large column density from the interstellar \NaI\
is required for a secure detection of outflow.  In
\fig{fig.index_mgb_vs_index_nad}, however, we see evidence that a
population of galaxies with outflows would be missed by a high
$W_0(\NaID) / W_0(\MgIb)$ selection scheme.  The figure is meant to
facilitate a comparison to the existing studies, in which the samples
are selected by a high level of star formation (i.e., a high infrared
luminosity).  As will be discussed later, outflows appear to outlive
starbursts, and the relic winds may present significant columns to be
detected in poststarburst or post--star-forming galaxies.

\subsection{Sample Selection}
\label{Sample Selection}

\begin{deluxetable}{cccc}
\tablecaption{Spectral Line Index Definitions}
\tablewidth{0pt}
\footnotesize
\tablehead{
\colhead{Line} &
\colhead{Blue continuum} &
\colhead{Line} &
\colhead{Red continuum}\nl
\colhead{} &
\colhead{(\AA)} &
\colhead{(\AA)} &
\colhead{(\AA)}
}
\startdata
\Hb   & 4836--4846 & 4846--4878 & 4878--4888\\
\MgIb & 5112--5142 & 5150--5200 & 5209--5239\\
\NaID & 5822--5842 & 5881--5910 & 5910--5930
\enddata
\label{tab.line_index}
\end{deluxetable}

We systematically search DEEP2 spectra in the EGS field for the
coverage of \NaID\ absorption line and continuum around it
(\tab{tab.line_index}).  An object is removed from our sample if the
redshifted \NaID\ spectral range is at least partially outside the
edge or falls on the gap between the blue and red CCD chips.  This
process reduces the sample to 2248 objects.  The spectral baseline of
DEEP2 observations restricts our sample to $0.11 < z <
0.54$.\footnote{Over the redshift range, the physical scale
corresponding to $1''$ slit varies from $2~\kpc$ to $9.5~\kpc$.}  For
each DEEP2 spectrum, {\hl the continuum \snr\ per pixel [\snrpp] in
the region} around \NaID\ is defined to be the median \snrpp\ computed
from the pixel flux values and their inverse variances registered
within the blue and red continua (\tab{tab.line_index}).  The main
selection cut is made at $\snrpp > 5$ around \NaID, reducing the
sample to 493 objects.  We also remove the objects whose \NaID\
feature is severely compromised by sky emission or atmospheric
absorption lines after visual inspection, reducing the sample to 431
objects.  The latter cut tends to remove objects at specific redshifts
where the redshifted \NaID\ overlaps with telluric features.

The particular choice of continuum \snr\ cut is a compromise between
the inclusion of more objects for better statistics and the
reliability of \NaID\ velocity measurements.  Given the limited
spectroscopic \snr, our desire to probe fainter objects is motivated
by the well-known, downsizing nature of star formation
\citep[e.g.,][]{cowi96} and the apparent correlation between the
presence of galactic-scale wind and the strength of star formation
\citep[e.g.,][]{mart05, rupk05p2}.  At low $z$, star formation is
expected in optically fainter, low surface brightness galaxies.  At
high $z$, increasingly brighter galaxies are host to star formation
yet become faint in their apparent brightness in the visible.  We also
expect that the high-\snr\ spectra are obtained from passive,
early-type galaxies, which are generally of high surface brightness.
All these effects conspire to make the galaxy population of interest
to be somewhat elusive in the optical selection used here.
Furthermore, the strength of \NaID\ absorption, both stellar and
interstellar, varies widely from galaxy to galaxy, and our ability to
detect an outflow depends on the strength of continuum as well as the
absorption feature.  It is therefore important to get some idea as to
what kind of \NaID\ outflow to which our measurements are sensitive in
this study.

\begin{figure}
\scalefigure
\plotone{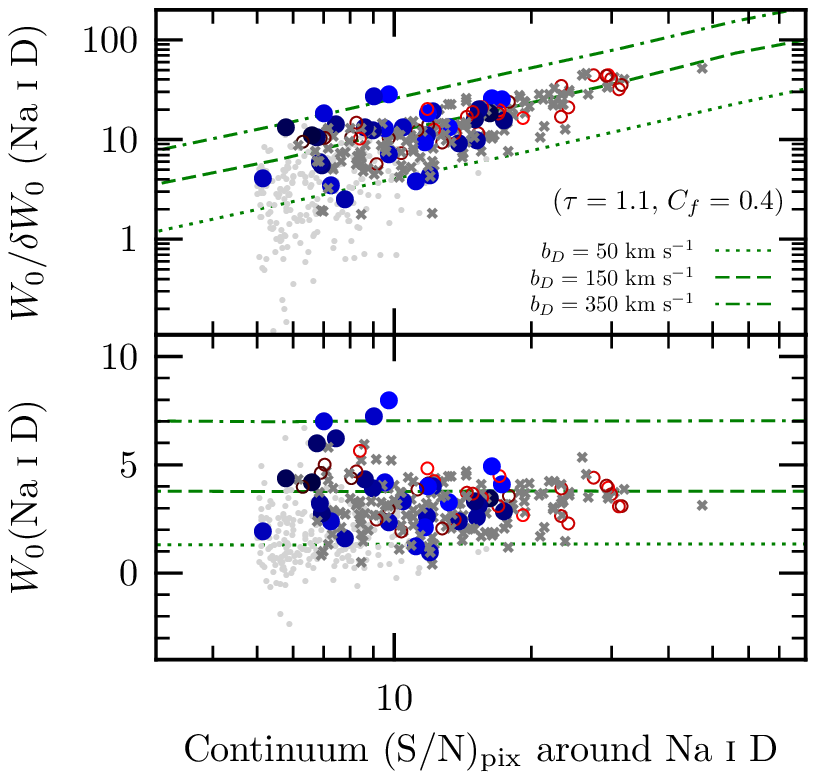}
\caption{ \snr\ (\emph{top}) and spectral line index
(\emph{bottom}) as a function of continuum \snrpp\ around \NaID.  The
points are denoted by \NaID\ kinematics: outflow (\emph{blue filled
circle}), systemic (\emph{gray cross}), inflow (\emph{red open
circle}), and low \snr\ (\emph{light gray dot}); see \sect{NaID
Velocity and Blueshift Probability} for the detail on \NaID\
kinematics.  Outflows/inflow data points are color-mapped with the
blueshift probability (\sect{NaID Velocity and Blueshift
Probability}), where bluer (redder) marks indicate stronger blueshift
(redshift) of \NaID\ absorption line.  The green lines are the loci of
\NaID\ indices measured in the simulated DEEP2 spectra assuming
fiducial \NaID\ absorption profiles expected from the fixed optical
depth and covering fraction ($\tau = 1.1$ and $\Cf = 0.4$) at three
different Doppler widths [$b_D = 50$ (dotted green line), $150$
(dashed green line), and $350~\km~\s^{-1}$ (dash-dotted green line)],
which are varied to simulate the effect of broadening in the
absorption profile (\emph{not} convolved with the instrumental
resolution of $\approx 42~\km~\s^{-1}$ here); see the Appendix for the
definitions of the variables.  The model with $b_D = 150~\km~\s^{-1}$
roughly reflects the mean \NaID\ property of LIRGs reported by
\citet{rupk05p2}; i.e., the long dashed line in the upper panel
indicates the significance of detection for a typical LIRG wind.  For
each model, a number of realizations are generated at a given \snr\
(i.e., a spectrum is degraded by Gaussian noise assuming that \snr) to
obtain the uncertainty in the spectral index.  (\emph{A color version of this figure is available in the online journal.})
}
\label{fig.snr_index_nad_vs_csnr}
\end{figure}

In \fig{fig.snr_index_nad_vs_csnr}, the distribution of spectral index
measurements as a function of continuum \snrpp\ around \NaID\ is shown
and demonstrates that we cannot obtain reliable \NaID\ velocity
measurements in most spectra (i.e., constitutes the low-\snr\ sample;
see \sect{Modeling NaID Absorption Lines} for detail) at $\snrpp \la
6.5$.  The detection limit as a function of \snrpp\ for a fiducial
LIRG wind indicate that we are reaching the $\sim 6\sigma$ detection
limit for a typical LIRG-type wind in our sample at that continuum
\snr\ level.  The distribution of low-\snr\ measurements suggests that
we lose the ability to detect and measure the kinematics of LIRG-like
outflows at $\snrpp \approx 5$.  Thus the selection cut at $\snrpp =
5$ seems justified, in terms of detecting LIRG-like winds at high
($\ga 5\sigma$) confidence.  \fig{fig.snr_index_nad_vs_csnr} shows
that the success rate of \NaID\ velocity measurement is a strong
function of continuum \snrpp.  Overall, there are 205 objects with
successful \NaID\ velocity measurements, and 226 objects without.  We
will describe what constitutes a ``successful'' (i.e., high \snr)
velocity measurement in \sect{NaID Velocity and Blueshift
Probability}.

\subsection{Modeling \NaID\ Absorption Lines}
\label{Modeling NaID Absorption Lines}

In principle, countless possible configurations of the geometry of
individual absorbers along a sightline give rise to an unlimited
variety of observed \NaID\ absorption line profiles; from numerical
simulations, several absorbers entrained in a starburst wind are
expected to lie along a single sight line (A. Fujita et al. 2008, in
preparation).  In practice, however, the spectral resolution and
moderate \snr\ limit our ability to study more than one component of
\NaID\ doublet in the DEEP2 spectra; multiple absorption components
would be seen blended even at a sufficient \snr.  Due to the
limitation, the physical quantities derived from a \NaID\ line profile
in general may not be uniquely determined.  Nonetheless, the modeling
of the absorption line should be physically motivated, and in this
respect we closely follow the method presented by \citet{rupk02,
rupk05p1}; readers are highly encouraged to find the detail of their
absorption line analysis method in those papers.  An absorption line
is modeled with the wavelength ($\lambda_c$) and the optical depth
($\tau_0$) at the line center, Doppler width ($b_D$), and the covering
fraction ($\Cf$) in a self-consistent manner.  Our confidence
intervals on \NaID\ velocity measurements, however, are obtained via
directly carrying out the Markov chain Monte Carlo (MCMC) analysis on
the observed spectra with the \citeauthor{rupk05p1}-type line-profile
modeling, rather than what \citet{rupk05p1} outlines.  In the
Appendix, we give a summary of the technique, the analysis method, and
the pipeline software developed for the task.

MCMC sampling generates a probability density for each model parameter
which allows us to visually inspect the quality of our measurements.
Except for a couple dozen high \snr\ spectra, the optical depths
cannot be constrained at all, i.e., the probability density for the
central optical depth $\tau_0$ usually ends up being distributed
roughly uniformly over the allowed range ($0 < \tau_0 < 999$) with a
slight enhancement toward the lowest optical depths.  In turn, the
highly saturated profile tends to let the covering fraction \Cf\ be
distributed near the level of minimum intensity of a \NaID\ profile.
This degeneracy between $\tau_0$ and $\Cf$ at a low \snr\ regime is a
well-understood property of the model profile employed by
\citeauthor{rupk05p1} and in this paper; extracting an optical depth
in general requires a line shape to be very well sampled.  The
distributions of \NaID\ central wavelengths $\lambda_c$ and Doppler
widths $b_D$, on the other hand, are relatively well behaving even at
lower \snr\s, where their probability densities roughly become
Gaussian.

One especially important caveat of our \NaID\ velocity measurement is
in order.  Our interest is in studying the interstellar gas
kinematics, so ideally stellar contributions to \NaID\ should be
removed via such a method as fitting template spectra generated from
population synthesis models \citep[e.g.,][]{trem04}.  The DEEP2
spectra are not rigorously fluxed, however, and we are unable to carry
out a similar procedure.  This is a substantial limitation in the
analysis of galaxy spectra of intermediate- and old-age stellar
populations, since their stellar absorption at \NaID\ becomes strong
(\fig{fig.index_mgb_vs_index_nad}; the stellar loci are from
\citet{delg05}).  The implication on our definition of outflow
velocity will be discussed in \sect{Definition of Outflow}.

We also do not take any special care of the nebular emission line
$\HeI~\lam~5876$, found $\sim 15~\ang$ blueward of \NaID\ in some
spectra.  The \HeI\ emission line can contaminate the high-velocity
tail of strong \NaID\ outflows in (U)LIRGs \citep[e.g.,][]{rupk05p1,
mart05}.  Upon visual inspection, however, few galaxies in our
sample with a good \NaID\ velocity measurement are found to have their
\NaID\ contaminated by the presence of the \HeI\ emission line.

\subsubsection{\NaID\ Velocity and Blueshift Probability}
\label{NaID Velocity and Blueshift Probability}

Each \NaID\ central wavelength $\lambda_c$ is converted to the \NaID\
velocity offset
\begin{equation}
v(\NaID) = c\frac{\lambda_c - \lambda_{\rm sys}}{\lambda_{\rm sys}} \ ,
\label{eqn.def_v}
\end{equation}
where $c$ is the speed of light and $\lambda_{\rm sys} = 5895.9243
(1+z)~\ang$ is the line center of \NaID\ shifted to the observed frame
using the cross-correlation redshift $z$.  (The redder line of the
doublet, $\NaI~\lam~5896$, is used as the reference line throughout.)
Using the above equation, the probability distribution of $v(\NaID)$
is directly obtained from the probability distribution of $\lambda_c$
from the MCMC sampling for each spectrum.  In order to take into
account the systemic redshift uncertainty, the probability
distribution is further convolved by a Gaussian kernel having a width
corresponding to the $1\sigma$ uncertainty in the cross-correlation
redshift (in the velocity space) for each object.  From each
probability distribution, the best estimate for \NaID\ velocity is
taken from the median, and the $68\%$ confidence interval is likewise
obtained.

\begin{figure}
\scalefigure
\plotone{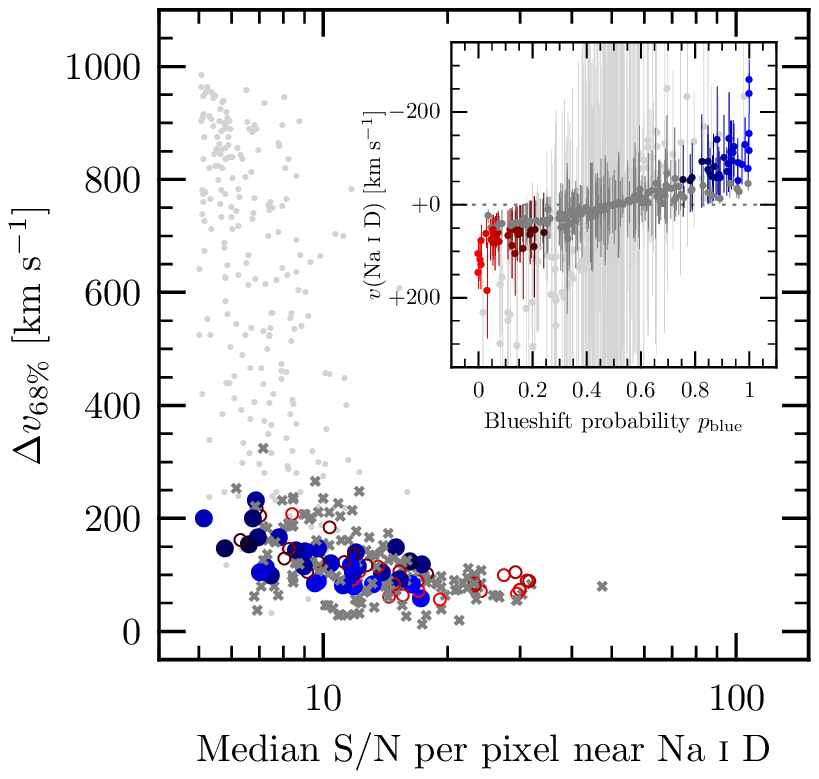}
\caption{ Distribution of \NaID\ velocity confidence intervals as a
function of continuum \snrpp\ around \NaID.  Here $\Delta v_{68\%}$ is
the difference of upper and lower $68\%$ confidence limits in \NaID\
velocity.  The symbols are as in \fig{fig.snr_index_nad_vs_csnr}.
\emph{Inset}: the \NaID\ velocity measurement with the error bar
indicating the $68\%$ confidence interval as a function of blueshift
probability \pblue.  Outflows and inflows from the high-\snr\ sample
are color-coded by blue and red, respectively, in gradient according
to \pblue\ (sect{NaID Velocity and Blueshift Probability}); those at
the systemic velocity are in gray.  The low-\snr\ velocity sample
(\sect{NaID Velocity and Blueshift Probability}) are in light gray.
High- and low-\snr\ velocity measurements are effectively separated in
the ways that they are distributed in these plots, indicating the
efficacy of the visual inspection scheme; see the text for detail.
(\emph{A color version of this figure is available in the online journal.})
}
\label{fig.dv_nad_vs_csnr_w_inset}
\end{figure}

For the purpose of measuring the \NaID\ velocity, the absorption
feature detected at a sufficient \snr\ for such a measurement
generally reveals itself as a normally distributed probability density
in $v(\NaID)$ well bounded within the parameter range, $v(\NaID) = \pm
700~\km~\s^{-1}$.\footnote{ While the \NaID\ outflow velocities
outside this interval have been reported in literature
\citep[e.g.,][]{rupk05p2, mart05}, the visual inspection of the
velocity measurements indicates that the range of $\pm
700~\km~\s^{-1}$ is sufficient for our sample, which does not
include ULIRGs (with signs of AGN activities) in which the most very
high velocity outflows are observed. }  At a lower \snr, a
probability distribution for $v(\NaID)$ exhibits high- and
low-velocity tails reaching the boundary values.  Hence, after visual
inspection, we divide our $v(\NaID)$ measurements into two classes
based on the behavior of $v(\NaID)$ probability distribution:
``high-\snr'' velocity sample ($N=205$) for which the distribution is well
within $\pm 700~\km~\s^{-1}$ and ``low-\snr'' velocity sample
($N=226$) for which the distribution either extends to the boundary or
is ill-behaving.  Visual inspection also guards against the sampling
results latching on to unwanted noise features, which usually show up
as an abnormal probability distribution function.  The MCMC
measurement pipeline also allows a fitted absorption-line profile to
be inspected for an interactively picked set of model parameters, so
the integrity of the fitting result has also been visually checked at
various points in the distributions of model parameters; see the
Appendix for detail.  \fig{fig.dv_nad_vs_csnr_w_inset} shows the
velocity width of confidence intervals ($\Delta v_{68\%}$; i.e., the
difference between the upper and lower $68\%$ confidence intervals) as
a function of \NaID\ continuum \snrpp.  Since $\Delta v_{68\%}$ is a
measure of the width of the probability distribution for $v(\NaID)$,
the figure shows that our strategy effectively distinguishes and
separates out high- and low-\snr\ velocity measurements, which tend to
appear at low and high $\Delta v_{68\%}$ regions, respectively.

\begin{figure}
\scalefigure
\plotone{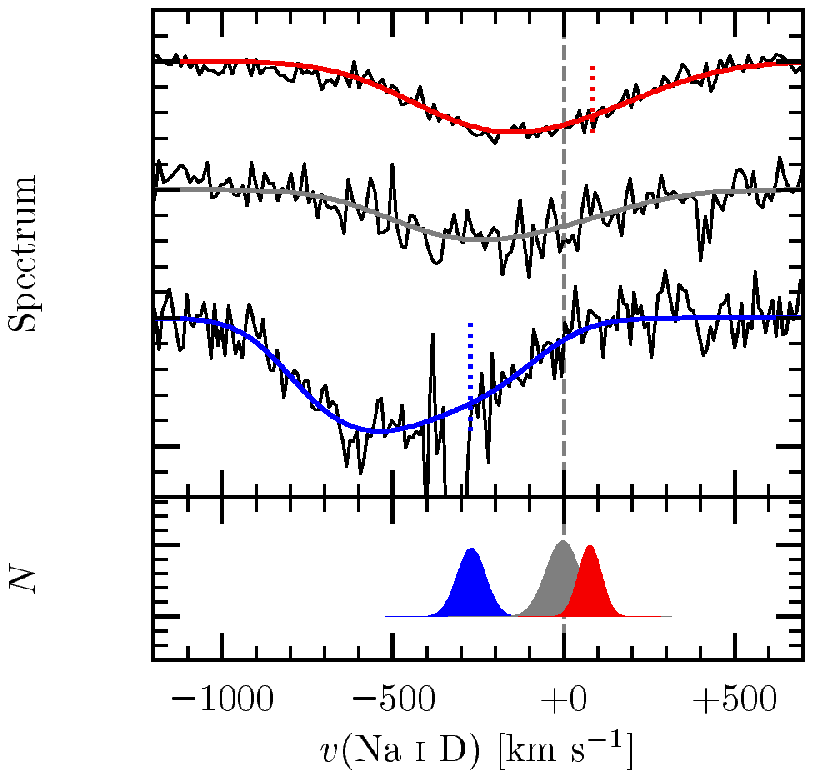}
\caption{ \emph{Top}: examples of continuum-normalized spectra around
\NaID\ (\emph{black line}) and the best-fit models for the blueshift
probability $\pblue \simeq 1$ (\emph{blue}), $0.5$ (\emph{gray}), and
$0.01$ (\emph{red}).  The vertical gray dashed line indicates the
systemic velocity.  The vertical dotted lines indicate the line center
of $\NaI~\lam~5896$ from the best-fit model for each spectrum.  The
$0~\km~\s^{-1}$ corresponds to the systemic velocity of the red line
of a blended doublet, and thus appears shifted toward red with respect
to the centroid of a line profile even if the doublet itself is at
systemic.  \emph{Bottom}: the probability distributions of \NaID\
velocity from the MCMC sampling; see \sect{NaID Velocity and Blueshift
Probability} for the detail on how the distributions are obtained.
For the object showing a very strong outflow (\emph{blue}), the
probability distribution is entirely found at $< 0~\km~\s^{-1}$, so
the blueshift probability \pblue\ approaches unity.  For a majority of
galaxies, the distributions are roughly symmetric about
$0~\km~\s^{-1}$, for which \pblue\ is about $0.5$ (\emph{black}).
(\emph{A color version of this figure is available in the online journal.})
}
\label{fig.outflow_prob_and_nad_spectra}
\end{figure}

Since we have robust probability density functions for $v(\NaID)$, it
is desirable to incorporate these into our definition of outflow
detection, rather than relying heavily on the best estimates for
$v(\NaID)$.  To do this, we introduce the \emph{blueshift probability}
\pblue\ for each velocity measurement, which is simply given by
\[
\pblue \equiv \int_{-\infty}^{0} dv \rho(v) \ ,
\]
where $\rho(v)$ is the probability distribution function obtained from
the MCMC sampling; recall that a blueshift yields a negative velocity
in Eq.~(\ref{eqn.def_v}), so the distribution function needs to be
integrated out to negative infinity to obtain the probability that a
\NaID\ is seen blueshifted in a given spectrum.  A few examples of
observed \NaID\ absorption spectra, along with the probability
distributions of \NaID\ velocity, are shown in
\fig{fig.outflow_prob_and_nad_spectra} for the cases of $\pblue \simeq
1$ (certainly an outflow), $0.5$ (likely at systemic), and $0.01$
(almost certainly ``inflow'').  The value of \pblue\ is a measure of
how likely that a ``real'' \NaID\ velocity is blueshifted from the
systemic velocity, given the result of MCMC sampling and the
uncertainty in the systemic redshift of a host galaxy.

\subsection{Definition of Outflow}
\label{Definition of Outflow}

To define what constitutes an outflow (``inflow'') detection for the
current analysis, we use a \NaID\ velocity cut of $v(\NaID) <
-50~\km~\s^{-1}$ ($> +50~\km~\s^{-1}$) and a blueshift probability cut
of $\pblue > 0.75$ ($< 0.25$).  Readers are cautioned that, based on
instrumental effects and data quality, precise definitions of outflows
necessarily vary in the literature.  The specific choice for the
velocity cutoff is primarily motivated by the typical $v(\NaID)$
uncertainty of $\sim 50~\km~\s^{-1}$
(\fig{fig.dv_nad_vs_csnr_w_inset}) and the visual inspection of
fitting results in relation to the values of $v(\NaID)$ and \pblue.
Although a blueshift probability, by construction, should be an
indicator of the likelihood of detecting an in/outflow at a desired
confidence level, the additional velocity cutoff is used to guard
against both the random and systematic errors in the redshift and the
velocity measurements.  For example, when the width of probability
distribution function is narrow, i.e., $\Delta v_{68\%}$ is small,
\pblue\ becomes more sensitive to the particular value of the systemic
velocity.  Upon visual inspection, we found that this often happened
when a \NaID\ doublet line shape was less blended, and therefore the
velocity moment could be determined more precisely.  In such a case,
an outflow as defined solely by a $v(\NaID)$ distribution function
looks dubious, having a tendency to be thrown off by an error in a
\emph{single} measurement of systemic redshift.  Ideally, our systemic
redshifts should also be scrutinized under rigorous MCMC analysis,
which we did not carry out.  For now, the additional velocity cut
adequately achieves the same goal.  The similar velocity cutoff was
employed by \citet{rupk05p2} to define outflow.

While employing a more stringent \pblue\ cut would yield a sample of
outflows detected with higher confidence, we then seriously have to
compromise our sample size.  In order to take advantage of the large
amount of data available in the AEGIS survey, our approach is to push
the limit of detections down to an acceptably low confidence level.
Hence our goal in this study is not to find individual cases of
outflows or inflows \emph{securely} but to give ourselves some
statistical power to characterize the general properties of host
galaxies.  This is a compromise that we opt to consciously make.

\begin{figure}
\scalefigure
\plotone{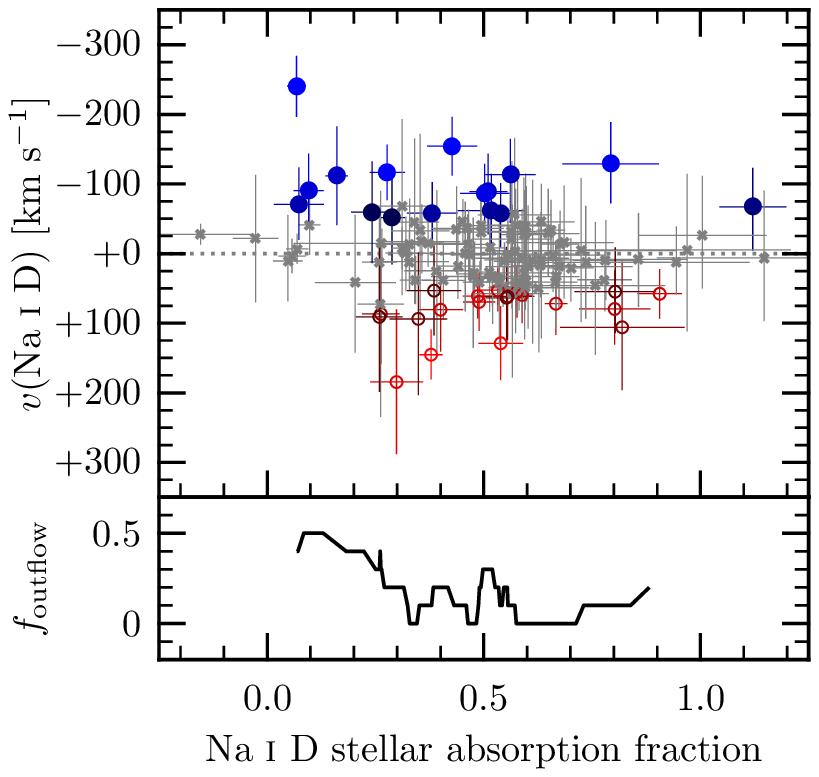}
\caption{ \emph{Top}: \NaID\ velocity as a function of stellar
fraction in the \NaID\ absorption index.  The stellar fraction is
given by $W_*(\NaID) / W_0(\NaID)$, where the stellar absorption
index is assumed to be $W_*(\NaID) = 0.4 W_*(\MgIb)$, so only the
objects with both \NaID\ and \MgIb\ measurements are plotted; see
the \MgIb-\NaID\ index plane (\fig{fig.index_mgb_vs_index_nad}).
For each object, the $1\sigma$ uncertainty in the stellar absorption
fraction is estimated from resampling its locations on the
\MgIb-\NaID\ index plane, drawn from a two-dimensional probability
distribution to match the known $W_0$ uncertainties in \NaID\ and
\MgIb.  The objects for which the uncertainty in their stellar
fraction is greater than $0.25$ are not plotted.  This removes three
outflows that appear to the left of stellar loci in
\fig{fig.index_mgb_vs_index_nad}; either \NaID\ or \MgIb\ is
contaminated by sky residuals in those objects, which cause their
stellar fraction uncertainties to be very high (greater than $0.7$). \emph{Bottom}: the fraction of objects hosting outflows as a
function of stellar fraction in the \NaID\ absorption index.  The
outflow fraction is computed at the sliding median of 10 adjacent
objects.  (\emph{A color version of this figure is available in the
online journal.})  }
\label{fig.v_nad_vs_sfrac_nad_w_outflow_frac}
\end{figure}

As emphasized earlier, one caveat of the \NaID\ absorption analysis in
this paper is our inability to remove the stellar contribution in the
\NaID\ absorption (\sect{Modeling NaID Absorption Lines}); a stellar
fraction can only be estimated indirectly from such an index as \MgIb\
whose strength is known to correlate well with that of \NaID.  Since
the presence of interstellar absorber(s) is a necessary condition for
outflows seen in absorption, some trend is expected to exist between
the outflow detection rate and the interstellar fraction of the total
\NaI\ column.  \fig{fig.v_nad_vs_sfrac_nad_w_outflow_frac} shows that
such a trend does exist, in which the objects with low \NaID\ stellar
absorption fractions are more likely to host outflows (that we can
detect).  Without the explicit removal of stellar absorption
components, however, the figure does not necessarily imply the paucity
of outflows in galaxies whose \NaID\ is dominated by the stellar
contribution.  It is important to note again that our census is only
sensitive to fairly strong outflows of the kind expected in LIRGs
(\fig{fig.snr_index_nad_vs_csnr}).  The kind of weak outflows observed
by \citet{schw04} in dwarf starbursts is certainly below the detection
level of the present survey.  Furthermore, since only one \NaID\
doublet component is fitted, the measured \NaID\ velocities are likely
the lower limits to the kinematic component with the highest outflow
velocity; each \NaID\ velocity is sensitive to the ``moment'' of
multiple absorption components, at least one of which is stellar in
origin and should exist at a systemic redshift.  In high-\snr\
spectroscopy of (U)LIRGs, \citet{rupk05p1} and \citet{mart05}, for
example, could fit more than one component of \NaID\ absorption in
some spectra, provided that line profiles are not overly ``smooth''
due to blending and instrumental smearing.
\fig{fig.v_nad_vs_sfrac_nad_w_outflow_frac} indicates that our
velocity measurements appear sensitive to outflows in \NaID\ when $\ga
50\%$ of the absorption equivalent width is interstellar in origin.
Given these limitations, attaching a physical meaning to a \NaID\
velocity would be misleading, since it is only the \emph{shift in the
moment} of the absorption profile, not the isolated absorption from
the outflowing/inflowing gas, to which our measurements are really
sensitive.  We thus avoid emphasis on the exact \NaID\ velocity
values.

\subsubsection{Reality of ``Inflows''}

The sensible cuts in the \NaID\ velocity and blueshift probability
\pblue\ give rise to a population of galaxies with ``inflows.''  While
it would not be surprising to see an inflow from a sightline through
an interacting system, ``inflows'' in our sample are seen mostly in
luminous, massive galaxies in the red sequence (\sect{Color Magnitude
Diagram}).  The presence of inflows in these objects is striking.
Given that these ``inflow'' detections are almost exclusively in
early-type galaxies presumably with relatively little interstellar
gas, we speculate whether factors other than interstellar absorption
can cause potential systematics.

First, the precision of cross-correlation redshifts is generally worse
with templates dominated by absorption lines, from which the redshifts
of galaxies with ``inflows'' are obtained.  It is also difficult to
visually inspect redshift systematics in absorption-dominated spectra
due to the lack of narrow, high-\snr\ features.  As mentioned in
\sect{Systemic Redshifts}, a systematic velocity difference of order
$10~\km~\s^{-1}$ also exists between cross-correlation and DEEP2
redshifts, although this is small compared to inflow velocities.  We
do, however, take the redshift uncertainty into account in our
definition of inflows (\sect{NaID Velocity and Blueshift
Probability}).  Second, it might be possible that some unaccounted
metal absorption features redward of \NaID\ may be causing
systematics, redshifting the single-component \NaID\ profile fit.  The
experimentation with synthesis spectra from \citet{bruz03} and
\citet{delg05} indicates that the degree of systematics would not be
as strong as observed, assuming that \NaID\ and nearby metal lines are
properly modeled.  Third, as a significant fraction of galaxies with
inflows appear to have lenticular morphology (Tremonti et al. 2007, private
communication), another possibility is that a distinct kinematic
component of \emph{stellar} motion might be detected in these systems,
perhaps from a faint disk.

While some of the above listed concerns are equally valid for
outflows, in following sections we show compelling evidence that the
detections of outflows are physically associated with star formation (or
AGNs/LINERs).  Such a clear physical connection cannot be made for
inflows at present, and a preliminary investigation to explore the
nature of the ``inflow'' population is underway.  In radio
ellipticals, neutral hydrogen is often seen in inflow, plausibly
feeding the nuclear activity \citep[e.g.,][]{vang89}; we do find some
consistency with this scenario in our inflows.  If these inflows are
indeed associated with such feeding of AGNs, our method could provide
an effective means of identifying the massive, early-type galaxies
going through the ``maintenance-mode'' of AGN feedback.  Nonetheless,
we mostly defer the discussions of inflows to future papers.

\subsection{On Selection Effects}

\emph{Dependence on \NaID\ absorption strength}.  Our ability to
measure a \NaID\ velocity depends primarily on the combination of the
absorption and the continuum strengths, which introduces selection
biases.  Our main selection cut on continuum \snr\ around \NaID\
feature favors luminous, high surface brightness objects, which tend
to be early-type galaxies on the red sequence (\sect{Color Magnitude
Diagram}).  Star-forming galaxies in the blue cloud are much fainter
at $z \la 0.5$ from which our sample is drawn.  In
\fig{fig.index_mgb_vs_index_nad}, we see that high $W_0(\NaID)$
objects tend to be old galaxies with a high stellar fraction or young
galaxies with a high interstellar fraction in their \NaID\ absorption.
Our survey lacks sensitivity to absorption lines in less luminous
star-forming galaxies, where much of star formation occurs at $z \la
0.5$.  Therefore, that a significant fraction of our outflow
detections is in red-sequence galaxies (\sect{Color Magnitude
Diagram}) is partially a selection effect.  Nonetheless, our sample
nicely complements the existing studies of \NaID\ outflows which have
focused on starburst galaxies.

\begin{figure}
\scalefigure
\plotone{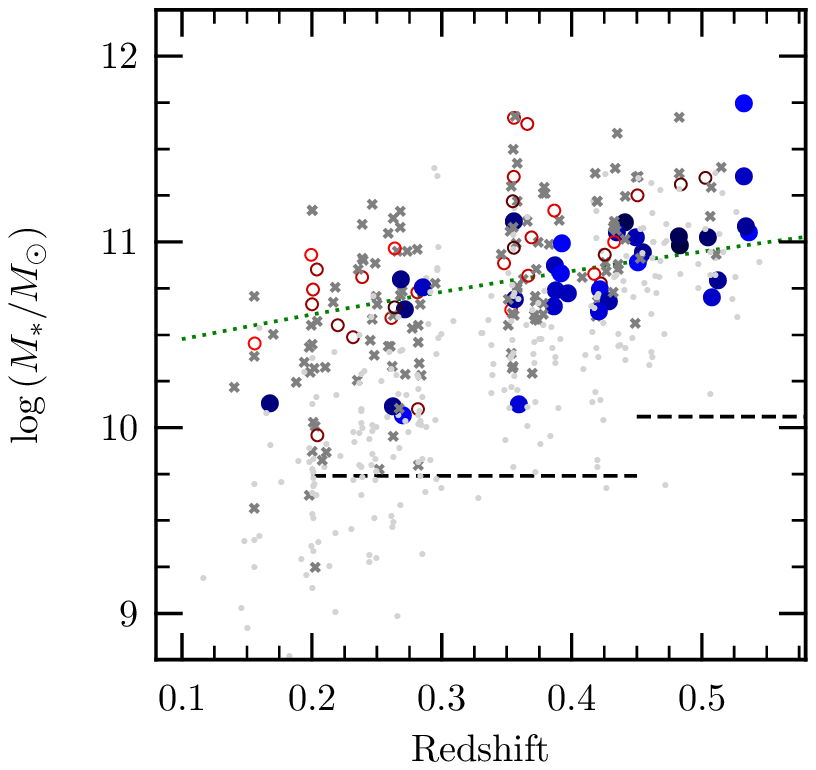}
\caption{ Stellar mass of galaxy as a function of redshift for the
sample in this study.  Each galaxy is coded by the \NaID\ kinematics
(\fig{fig.dv_nad_vs_csnr_w_inset}): outflow (\emph{blue filled
circle}), systemic (\emph{gray cross}), inflow (\emph{red open
circle}), and low \snr\ (\emph{light gray dot}).  The horizontal
dashed lines roughly indicate the $95\%$ completeness limit in stellar
mass computed at the centers of redshift intervals for $0.2 < z <
0.45$ and $0.45 < z < 0.7$ from \citet{noes07a}.  The green dotted
curve indicates the ``quenching mass limit'' as presented in
\citet{bund07}.
(\emph{A color version of this figure is available in the online journal.})
}
\label{figs.mass_star_vs_z}
\end{figure}

\emph{Redshift dependence}.  Our sample consists of objects
distributed over a fairly large redshift interval of $0.11 < z <
0.54$, and the analysis suffers from the problem typically encountered
by a flux-limited survey.  First, the observation becomes increasingly
insensitive to fainter populations at higher redshift.  Second, given
a fixed opening angle, the survey volume tends to be smaller at lower
$z$, and the sample variance (i.e., cosmic variance as galaxy
enthusiasts like to say) becomes a serious issue; the survey field of
view changes by a factor of $\approx 10$ in physical area over the
redshift interval.  Third, the lookback time difference of several
\Gyr\ means that the galaxies experience a significant evolution over
that redshift interval, so the lower- and higher-$z$ populations
characterized by one ``fixed'' physical property may not be of the
same kind when looked at in view of other properties.

Some insights into these effects are gained from
\fig{figs.mass_star_vs_z}.  Apparently, the high-\snr\ sample
(\sect{NaID Velocity and Blueshift Probability}) does not come close
to the completeness limit in stellar mass of the parent AEGIS survey,
except at the lowest redshifts.  The lower sampling rate of high-mass
galaxies (e.g., $\log{\left(M_*/M_\odot\right)} \ga 11$) at low
redshift may be a result of the smaller survey volume; that is,
luminous early-type galaxies tend to be highly clustered
\citep[e.g.,][]{coil07}, and a smaller number of overdense regions
with bright ellipticals fall in the field of view toward lower
redshift.  The effect of the changing physical aperture size, from
$2~\kpc$ to $9.5~\kpc$ corresponding to the $1''$ slit, on the
detectability of outflow is difficult to assess.  The work by
\citet{mart06} on the spatially resolved \NaID\ outflows in local
ULIRGs indicates that the extended blueshifted interstellar absorption
is found over the scale of $\ga 15~\kpc$, in which case a significant
column should remain available within the regions covered by the slit
over the redshift range covered in this study.

\section{Host Galaxies of Outflows}
\label{Host Galaxies of Outflows}

\subsection{Trends with Star Formation}
\label{Trends with Star Formation}

\begin{deluxetable}{lccc}
\tablecaption{Detection Rates for \NaID\ Outflows Selected by \Lir}
\tablewidth{0pt}
\footnotesize
\tablehead{
\colhead{Criterion} &
\colhead{$N_{\rm subsample}$} &
\colhead{$N_{\rm outflow}$} &
\colhead{Detection Rate}
}
\startdata
Within MIPS coverage          & 169 &  24 & $0.14 \pm 0.03$ \\
$\log{(\Lir/L_\odot)}  > 11$  &  21 &   8 & $0.38 \pm 0.11$ \\
$\log{(\Lir/L_\odot)} \le 11$ &  64 &   5 & $0.08 \pm 0.03$ \\
No $24\mu$ detection         &  84 &  11 & $0.13 \pm 0.04$
\enddata

\tablecomments{ The conventional cut for LIRGs is
$\log{(\Lir/L_\odot)} > 11$.  Only the objects with high-\snr\ \NaID\
velocities are included for the calculation of detection rates; i.e.,
the objects in the low-\snr\ sample (\sect{NaID Velocity and Blueshift
Probability}) are treated as if they are not detected.  The
uncertainties are estimated from binomial statistics.  }

\label{tab.detection_rates}
\end{deluxetable}

\begin{figure}
\scalefigure
\plotone{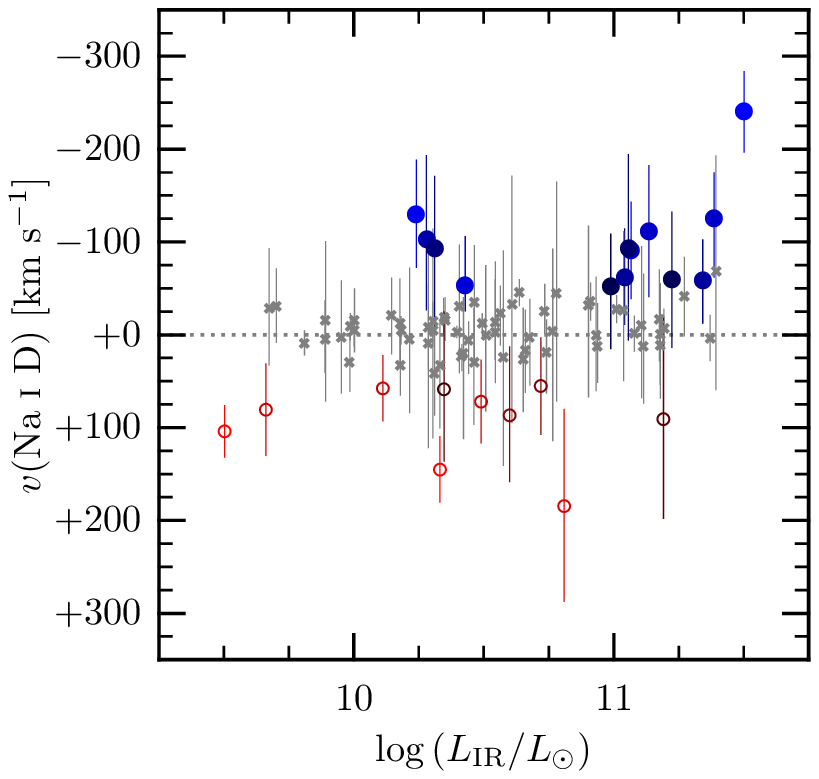}
\caption{ \NaID\ velocity as a function of infrared luminosity for the
objects in the high-\snr\ sample (\sect{NaID Velocity and Blueshift
Probability}).  The figure includes only the objects that are
covered by the \emph{Spitzer}/MIPS
observations and are detected above the flux limit ($S_{24} \sim 83
\mu~\Jy$) to yield the measurements of \Lir.  The symbols are as in
\fig{fig.snr_index_nad_vs_csnr}.
(\emph{A color version of this figure is available in the online journal.})
}
\label{fig.v_nad_vs_lir}
\end{figure}

Since the recent advance in the detailed knowledge of starburst-driven
galactic winds have been gained through the studies of local LIRGs
\citep[e.g.,][]{heck00, rupk02, rupk05p1, rupk05p2, mart05}, the
\NaID\ velocity measurements as a function of infrared luminosities
$\Lir$ naturally provide a starting point for comparison.  Since for
star-forming galaxies the infrared is dominated by the thermal dust
emission of reprocessed starlight from hot, young massive stars, a
tight correlation exists between \Lir\ and SFR in dusty star-forming
galaxies \citep{kenn98}.  For a subset of our sample, the far-infrared
photometry from \emph{Spitzer}/Multiband Imaging Photometer
\citep[MIPS;][]{riek04} is used to derive the total infrared
luminosity \Lir, following \citet{lefl05} and using the \citet{char01}
spectral energy distribution (SED) templates.

In \fig{fig.v_nad_vs_lir}, a clear tendency is observed for high
\Lir\ objects to host outflows.  Indeed, a majority of outflows are
found in LIRGs ($\Lir > 10^{11}~L_\odot$) in the figure.
\tab{tab.detection_rates} presents the outflow detection rates for the
subsampling based on \Lir.  The outflow detection rate of $38 \% \pm
11 \%$ for LIRGs is similar to those reported by low-$z$ surveys of
infrared-selected galaxies; e.g., $42 \% \pm 8 \%$ in \citet{rupk05p2}
and $32 \% \pm 12 \%$ in \citet{heck00}.  Hence we confirm the results
reported by others that the detection rate of outflow in
infrared-selected galaxies correlates well with their infrared
luminosity.  We note, however, that the exact values for the detection
rates, especially among the galaxies with lower \Lir, may not be
robust against selection effects and incompleteness, since our primary
selection cut is made by the strength of continuum around
\NaID\ (\sect{Sample Selection}).  There are no ULIRGs
[$\log{(\Lir/L_\odot)} > 12$] in our sample, which is
reasonable given the expected number of ULIRGs in the survey
volume is very small ($\la 10$), assuming redshift-dependent
luminosity functions of infrared sources \citep{lefl05} and the
completeness of the DEEP2 survey.

\begin{figure}
\scalefigure
\plotone{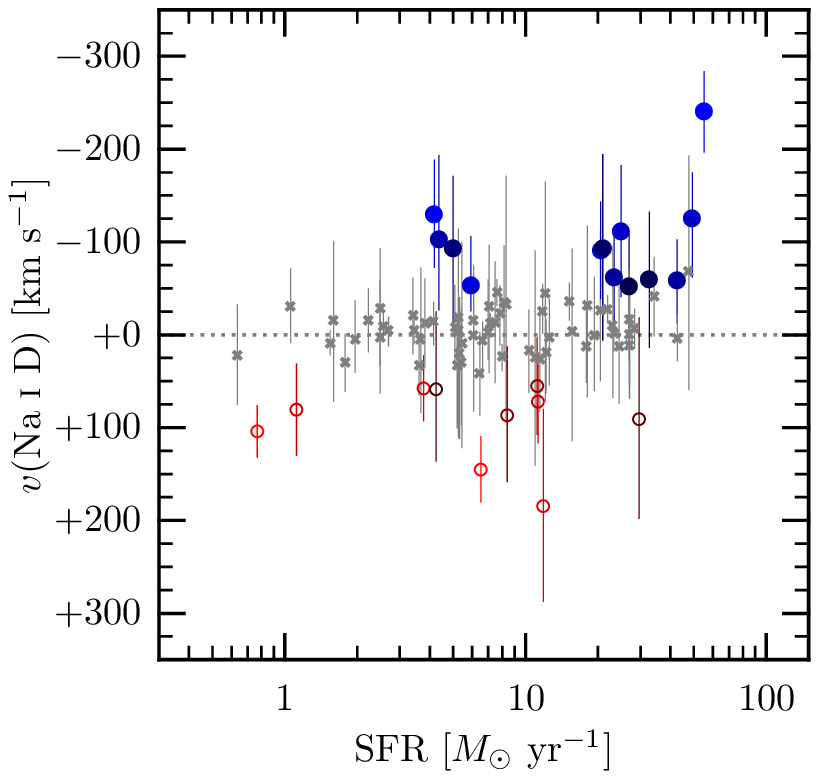}
\caption{ \NaID\ velocity as a function of total SFR, i.e., the sum of
SFRs from the infrared emission and optical emission lines
(\sect{Trends with Star Formation}).  The symbols are as in
\fig{fig.snr_index_nad_vs_csnr}.  The SFR corresponding to that of
LIRG ($\Lir > 10^{11}~L_\odot$) is $\approx
17~\dot{M}_\odot~\yr^{-1}$; see \fig{fig.v_nad_vs_lir} for comparison.
(\emph{A color version of this figure is available in the online journal.})
}
\label{fig.v_nad_vs_sfr_tot}
\end{figure}

In \fig{fig.v_nad_vs_sfr_tot}, the \NaID\ velocity as a function of
the total SFR (i.e., the sum of SFRs derived from infrared SED and
optical emission lines; see \citet{noes07a} for details) is shown.
The similar dependence of $v(\NaID)$ on \Lir\ and the total SFR is
expected due to the tight correlation between \Lir\ and the total SFR,
which is often dominated by the infrared contribution.  Again, a
majority of outflows are seen in galaxies with SFR $\ga
20~M_\odot~\yr^{-1}$, which correspond to the amount of SFR expected
in LIRGs.  As a caveat, it should be mentioned that some of the
infrared luminosity might be from (obscured) infrared-bright AGN, and
not from star formation.  We lack the diagnostics to distinguish star
formation from AGN, but the simple relation between \Lir\ and SFR may
not hold for the cases in which AGN contributions to the infrared
luminosities are high.  {\hl Local studies of LIRGs suggest that
strong AGN contamination may be small, though low-luminosity AGN
could be common \citep[e.g.,][]{veil95}.}

\begin{figure}
\scalefigure
\plotone{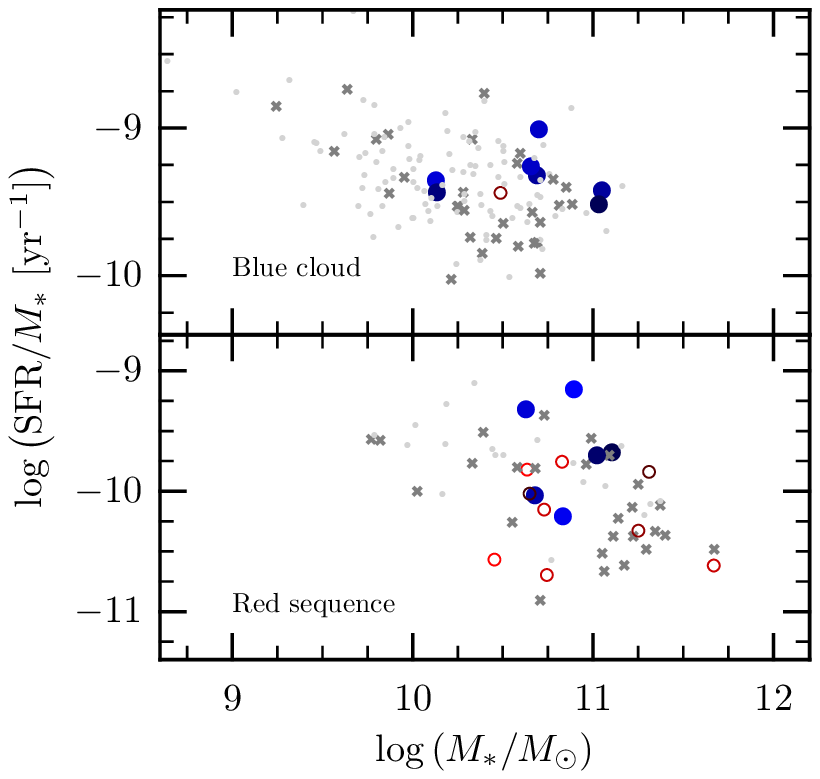}
\caption{ Specific SFR as a function of stellar mass for blue-cloud
(\emph{top}) and red-sequence (\emph{bottom}) galaxies; see
\sect{Color Magnitude Diagram} for the color cut scheme.  The symbols
are as in \fig{fig.snr_index_nad_vs_csnr}.  The figure is comparable
to Fig.~1 of \citet{noes07b}, except that red AGN/LINER candidates are
not removed.  An SFR estimate from optical line emission suffers
significantly from AGN/LINER contamination \citep{yan06, wein07};
therefore, their specific SFRs are likely overestimates here,
especially for the red-sequence galaxies.
(\emph{A color version of this figure is available in the online journal.})
}
\label{fig.ssfr_vs_mass_star}
\end{figure}

Since the SFR scales almost proportionally with galaxy mass
\citep[e.g.,][]{brin04}, its inference with respect to the specific
amount of star-forming activity may be misleading.  The birthrate
parameter, $b \equiv {\rm SFR}/\langle{\rm SFR}\rangle$
\citep{kenn98}, is a better indicator which takes into account the
star formation history yet is difficult to estimate due to its
dependence on the timescale over which the galaxy has been forming
stars as well as on the fraction of gas recycled in the past.  It is
still desirable to remove the first-order effect of galaxy mass, since
our sample is drawn over the redshift interval where downsizing
affects star formation particularly strongly.  Specific SFR, defined
as ${\rm SFR} / M_*$ (i.e., SFR per unit stellar mass) is easier to
compute and a reasonable proxy for birthrate parameter.
\fig{fig.ssfr_vs_mass_star} shows how the specific SFRs are
distributed as a function of galaxy stellar mass $M_*$, obtained from
fitting SEDs to optical and near-infrared photometry \citep{bund06}.
For the red-sequence galaxies (\fig{fig.ub_0_vs_m_b}), the specific
SFRs may be overestimated due to AGN/LINER contamination \citep{yan06,
wein07}.  Focusing on the blue-cloud galaxies, an interesting
feature is that, at a fixed mass, a majority of the objects with
outflows are seen at the upper envelope of the distribution of
specific SFRs.  The analysis of the parent AEGIS sample by
\citet{noes07a, noes07b} indicates that, in the redshift interval
$0.11 \la z \la 0.55$ from which our sample is drawn, $M_* \sim
10^{11}~M_\odot$ is roughly where star formation is seen to be
quenched in the most massive galaxies.  In the full \citet{noes07b}
sample, the distribution of star-forming galaxies is fairly tight; the
loci in \fig{fig.ssfr_vs_mass_star} suggest that host galaxies of
outflows are undergoing an enhanced episode of star formation,
relative to the star-forming galaxies without outflows at the same
epoch.

It is emphasized that star formation may not be the only source of
optical/infrared emission in the presence of an AGN.  We see that
quite a few red-sequence galaxies with \emph{inflows} show up in the
figures involving SFRs presented in this section.  As discussed in the
next section, these objects show little evidence of ongoing star
formation in UV diagnostics, and most are either quiescent or show
Seyfert/LINER-like excitation in their $\fNII/\Ha$ emission line
ratios (\sect{Star Formation versus AGN}).  Hence inflows in general
are more likely to be associated with AGN/LINERs than star formation.

\subsection{Color Magnitude Diagram}
\label{Color Magnitude Diagram}

Although optical photometry only tells an incomplete history, dividing
galaxies into blue and red populations by optical colors still is very
useful for capturing the essence of galaxy evolution with a bird's-eye
view, since the bimodality of galaxy color distribution is among the
most prominent features that persist over wide ranges of parameters,
such as galaxy mass, luminosity, environment, and redshift
\citep[e.g.,][]{bald04, balo04, will06, coop07a}.  Over the past few
decades, numerous studies of local galaxies have shown that the ``red
sequence'' is generally populated by red, dead, passively evolving
galaxies with old stellar populations, while the ``blue cloud'' is
populated by blue, actively star-forming galaxies.  These two densely
populated regions are divided by a sparsely populated area, sometimes
called the ``green valley''; see \citet{fabe07} and references therein
for a comprehensive account.

\begin{figure}
\scalefigure
\plotone{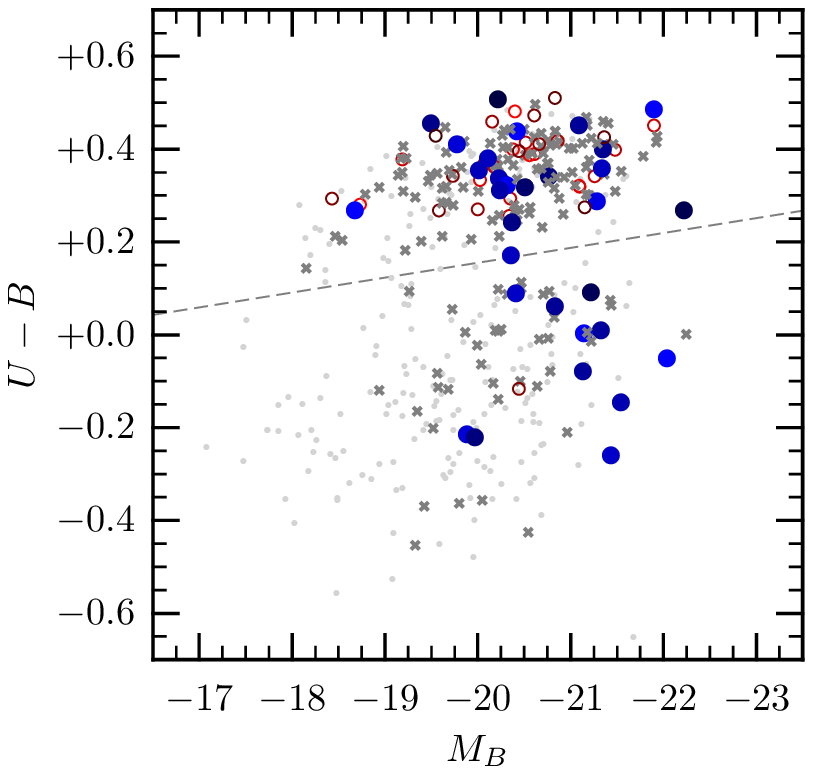}
\caption{ Rest-frame $(U-B)$-$M_B$ color-magnitude diagram.  The
photometry, on the Vega system, is corrected for Galactic extinction
but not for the internal extinction of galaxies; see \citet{will06}
for detail.  The symbols are as in \fig{fig.snr_index_nad_vs_csnr}.
The dashed line indicates the $U-B$ color cut used to divide
red-sequence and blue-cloud galaxies; see \citet{fabe07}.  [\emph{See
the electronic edition of the paper for a color version of this
figure.}]}
\label{fig.ub_0_vs_m_b}
\end{figure}

\fig{fig.ub_0_vs_m_b} presents the rest-frame $(U-B, M_B)$
color-magnitude diagram, in which the host galaxies of outflows are
seen in relation to the red sequence and the blue cloud.  The catalog
of rest-frame photometry was constructed as described in
\citet{will06}.\footnote{ The optical photometry are on the Vega
system.  See \citet{will06} for the conversion procedure between
Vega and AB magnitudes. }  Given that the low-$z$ red sequence
galaxies are characterized as inactive and quiescent, that a majority
of outflows are found in red galaxies comes as a surprise; among 32
outflows, 21 of them are found on the red sequence.
Non--star-forming, quiescent galaxies are red in the rest-frame
optical due to their intrinsic SEDs; the rest-frame $U-B$ in
particular brackets the $4000~\ang$ feature, which is sensitive to the
``break'' caused by the Balmer limit, as well as the high metal
opacity in the cool stellar atmosphere of old stars.  Nonetheless,
high dust reddening can push some star-forming galaxies into the red
sequence, especially at higher redshift \citep[e.g.,][]{wein05,
bell05}.  The rest-frame photometry in \fig{fig.ub_0_vs_m_b} are not
corrected for the internal extinction of galaxies, so the
``contamination'' from dust-reddened star-forming galaxies can happen
though expected to be small at $z < 0.5$.  We also reiterate that, due
to the fixed continuum \snr\ cut, our sample is naturally biased for
red-sequence galaxies, which tend to have high surface brightness
(\sect{Sample Selection}).  {\hl This selection bias due to the
continuum strength can be inferred from the difference in the
outflow detection rates for red and blue subpopulations.  Among the
full sample ($N = 431$), the outflow detection rates are $11/205 =
0.05 \pm 0.02$ and $21/226 = 0.09 \pm 0.02$ for blue and red
objects, while they are $11/51 = 0.22 \pm 0.06$ and $21/154 = 0.14
\pm 0.03$ among the high-\snr\ sample ($N = 205$).  }

\begin{figure}
\scalefigure
\plotone{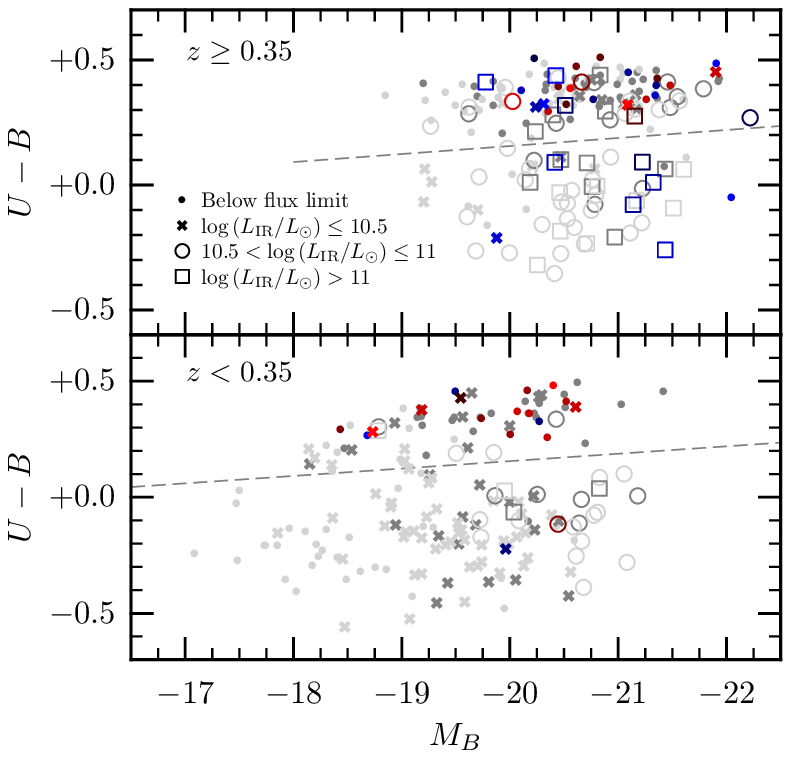}
\caption{ Rest-frame $(U-B)$-$M_B$ color-magnitude diagram for objects
at $z \ge 0.35$ (\emph{top}) and $z < 0.35$ (\emph{bottom}).  The gray
dashed lines divide red and blue galaxies (\fig{fig.ub_0_vs_m_b}).
The symbol indicates the infrared luminosity \Lir: below flux limit
(\emph{dot}), $\log{(\Lir/L_\odot)} \le 10.5$ (\emph{cross}), $10.5 <
\log{(\Lir/L_\odot)} \le 11$ (\emph{open circle}), and
$\log{(\Lir/L_\odot)} > 11$ (\emph{open square}).  The color scheme
for the symbols is similar to \fig{fig.snr_index_nad_vs_csnr} and
indicates \NaID\ kinematics: outflow (\emph{blue}), inflow
(\emph{red}), systemic (\emph{gray}), and low \snr\ (\emph{light
gray}).  There are roughly equal number of objects above and below the
redshift cut at $z = 0.35$.  [\emph{See the electronic edition of the
paper for a color version of this figure.}]}
\label{fig.ub_0_vs_m_b_w_lir}
\end{figure}

\fig{fig.ub_0_vs_m_b_w_lir} shows the color-magnitude diagram in which
the objects are denoted by their infrared luminosity and offers
evidence that star-forming galaxies do reside in the red
sequence, though relatively small in fraction.  The sample is
divided at $z = 0.35$ into low- and high-redshift subsamples, which
makes clear that the infrared-luminous galaxy fraction is much greater
at higher redshift, consistent with lookback-time studies of
star-forming, infrared-luminous galaxies \citep[e.g.,][]{bell05,
lefl05, noes07a}.  Outflows are mostly observed in $z > 0.35$
objects.  A comparison of the top and bottom panels shows quite
strikingly that the $z > 0.35$ red sequence has a substantial number
of infrared-luminous galaxies, some of which are LIRGs, which is
absent in the $z < 0.35$ red sequence.  The red-sequence outflows,
however, are also found in the objects without significant $24~\mu{\rm
m}$ flux.

\begin{figure}
\scalefigure
\plotone{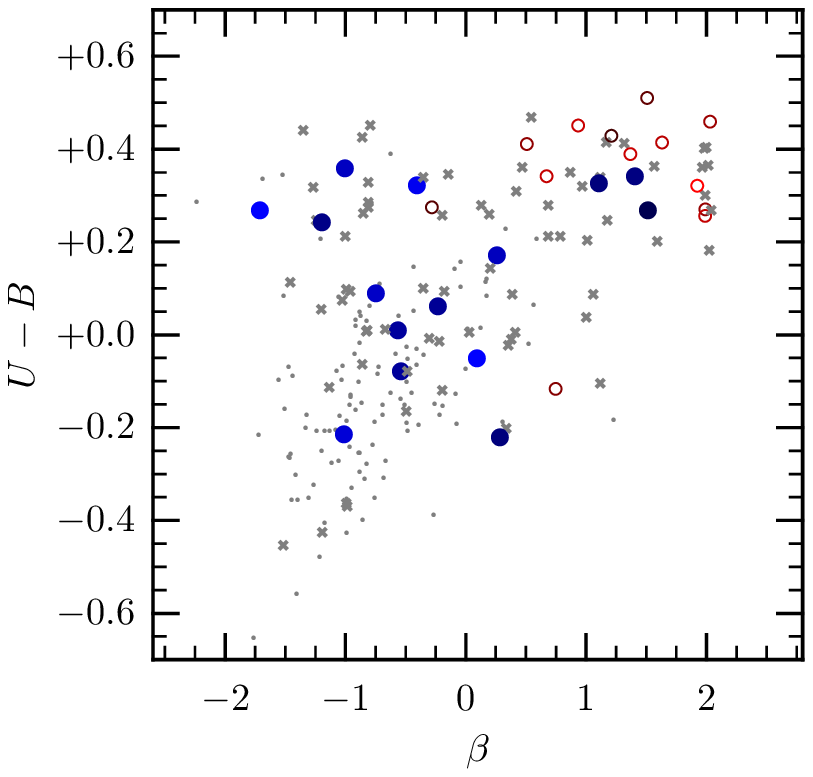}
\caption{ Rest-frame $U-B$ color as a function of the UV spectral
slope $\beta$ for a subset of objects with GALEX photometry.  The
$\beta$ follows the definition given by \citet{seib05}, $f_\lambda
\propto \lambda^\beta$; star-forming galaxies have smaller values
of $\beta$ due to their rising continua toward shorter wavelengths.
The symbols are as in \fig{fig.snr_index_nad_vs_csnr}.  A majority of
blue-sequence galaxies are classified as low-\snr; they are bright in
UV but fainter in the visible, making it difficult to obtain their
high-\snr\ spectra.
(\emph{A color version of this figure is available in the online journal.})
}
\label{fig.ub_0_vs_beta}
\end{figure}

\begin{figure}
\scalefigure
\plotone{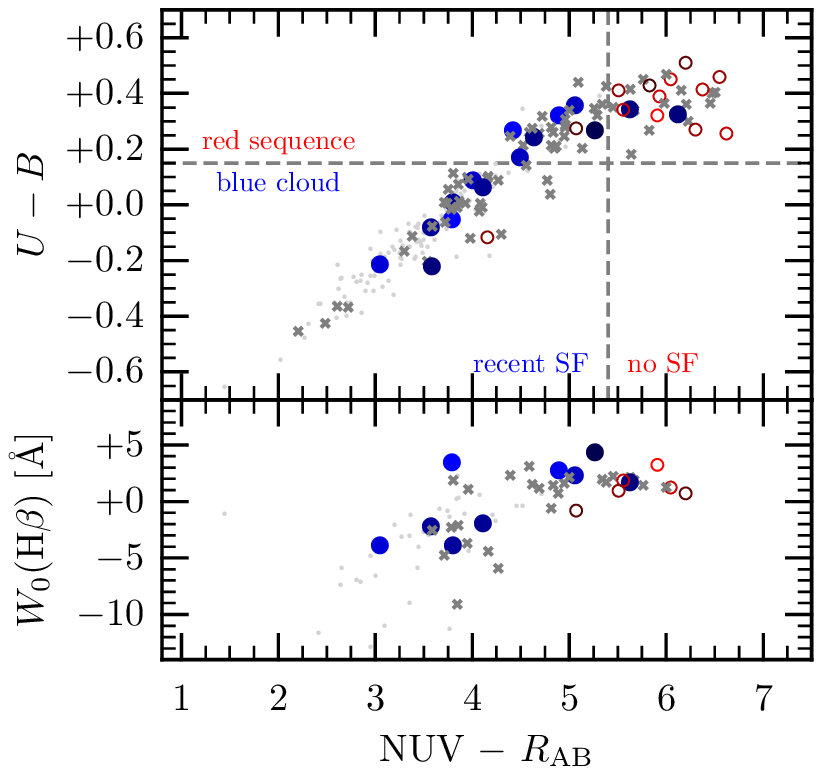}
\caption{ \emph{Top}: rest-frame $U-B$ color as a function of
rest-frame $\NUV - R_{\rm AB}$ color.  The horizontal dashed line at
$U-B = 0.15$ roughly divides red-sequence and blue-cloud galaxies
(\fig{fig.ub_0_vs_m_b}).  The vertical dashed line at $\NUV - R_{\rm
AB} = 5.4$ follows the delineation of star-forming and
non--star-forming galaxies employed by \citet{scha06}, who used $\NUV
- r$; $r$ is roughly comparable to $R_{\rm AB}$.  The symbols are as
in \fig{fig.snr_index_nad_vs_csnr}.  \emph{Bottom}: \Hb\ spectral line
index as a function of rest-frame $\NUV - R_{\rm AB}$ color for a
subset of objects with the \Hb\ coverage.  The strength of \Hb\
absorption [i.e., $W_0(\Hb) \ga 0~\ang$] is sensitive to a starburst
occurred within $\sim 10^7~\Myr$--$2~\Gyr$ ago (see the inset of
\fig{fig.v_nad_vs_index_hb}).  A negative $W_0(\Hb)$ indicates the
presence of emission at \Hb, i.e., $W_0$ is an estimate of the sum of
emission and absorption equivalent widths (\sect{Spectral Line
Indices}).
(\emph{A color version of this figure is available in the online journal.})
}
\label{fig.ub_0_and_index_hb_vs_m_nuv}
\end{figure}

While the narrow spectral baseline of DEEP2 makes the Balmer decrement
unavailable, the UV spectral slopes $\beta$ from the \emph{Galaxy
Evolution Explorer} (GALEX) photometry are measured for a subset of
our sample.\footnote{The images were processed using ver. 4.1 of
\emph{GALEX} pipeline; see http://www.galex.caltech.edu.}  The UV spectral
slope is generally a good measure of dust reddening, since it is a
direct measure of continuum slope in ultraviolet, where dust optical
depth is particularly high.  \fig{fig.ub_0_vs_beta} shows that some
red galaxies with outflows have $\beta \la 0$, comparable to typical
blue star-forming galaxies.  None of these with $24~\mu{\rm m}$
detection, however, are LIRGs.  Another look into their identity is
given by the top panel of \fig{fig.ub_0_and_index_hb_vs_m_nuv}, where
the distribution of galaxies in their rest-frame $U-B$ color is
compared to that of the rest-frame $\NUV - R_{\rm AB}$ color.  The
UV-optical color is much more sensitive to a small fraction of young
stellar populations and in turn the current as well as recent
star-forming activity than optical colors alone.  The particular $\NUV
- R_{\rm AB}$ cut delineating star formation from no star formation is
from \citet{scha06}, who used $\NUV - r$ colors; SDSS $r$ is close to
CFHT $R_{\rm AB}$.  In this empirical color cut, early-type galaxies
with the strongest UV upturn observed locally should not contaminate
the star-forming classification; this makes inevitable that some
objects in the ``no star formation'' region may actually be
star forming.  Based on the cut, \citeauthor{scha06} found that $\sim
30\%$ of visually classified $z < 0.1$ early-type galaxies brighter
than $M_r = -21.5$ showed signs of recent star formation.  Strikingly,
most optically red galaxies with outflows have $\NUV - R_{\rm AB} <
5.4$, suggestive of recent star formation.

These results give rise to an interpretation that the arrival on the
red sequence of the galaxies with outflows happened only recently;
i.e., red sequence outflows are found predominantly in
post--star-forming galaxies with detectable residual star formation or
dusty star-forming galaxies with high infrared emission.  The redness
in their visible colors might also arise partly from the presence of
dust; the strong correlation between reddening $E(B-V)$ and the
equivalent width of low-ionization absorption lines are often reported
\citep[e.g.,][]{armu89, veil95, heck00}.  Poststarburst galaxies,
identified spectroscopically by their strong Balmer absorption, are
generally found to be dusty as well \citep[e.g.,][]{pogg00, pogg01,
balo05, sato06}.  The appearance of the post--star-forming phase also
fits well with the scenario that it is observed when star-forming
galaxies in the blue cloud make the transition to the red-sequence
after some mechanism triggered an enhanced episode of star formation,
which then gets shut off.  The $\NUV - R_{\rm AB}$ color alone,
however, does not rule out low-level star formation in these galaxies,
since whether a galaxy goes through poststarburst depends on the
fraction of mass that has formed in the most recent star formation
event as well as its timescale.  It is also worth noting the clear
separation between outflows and inflows in $\NUV - R_{\rm AB}$; most
inflows are observed in red galaxies in which little or no star
formation is detected.

\subsection{Evidence for Poststarburst}
\label{Evidence for Poststarburst}

\begin{figure}
\scalefigure
\plotone{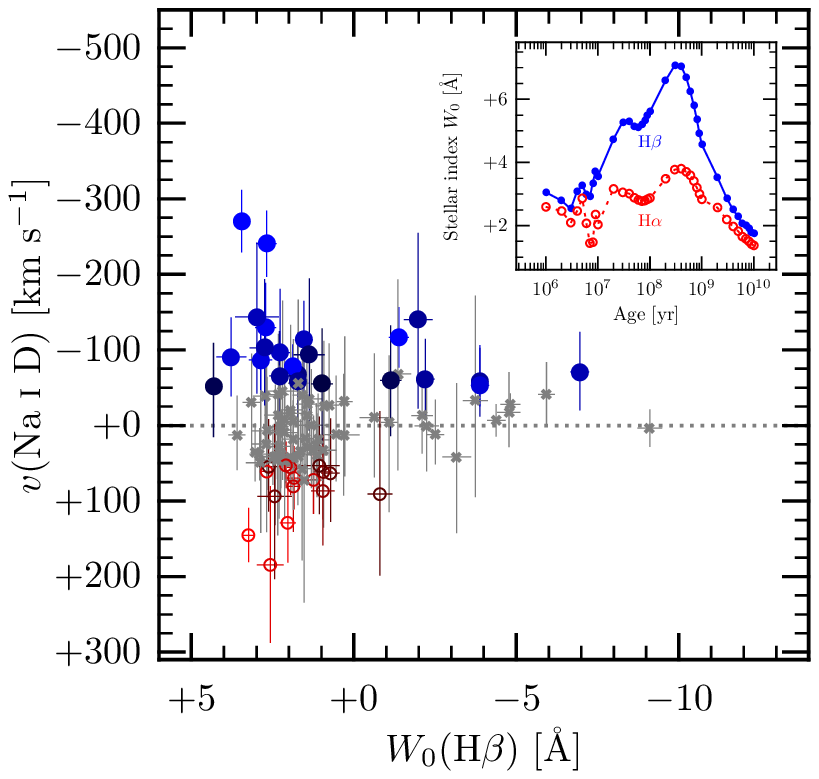}
\caption{ \NaID\ velocity as a function of \Hb\ spectral line index.
The symbols are as in \fig{fig.snr_index_nad_vs_csnr}.  Since the \Hb\
index measures the sum of absorption and emission lines, the \Hb\
index becomes a lower limit to the strength of \Hb\ absorption
equivalent width, if emission filling is significant.  In turn,
$W_0(\Hb) \la 0~\ang$ indicates that the emission line flux is greater
than that of absorption line. See \sect{Spectral Line Indices} for
detail on the definition of the spectral line index.  \emph{Inset}:
the stellar spectral line index at \Ha\ (\emph{red dashed line}) and
\Hb\ (\emph{blue solid line}) as a function of stellar ages.  The
indices are measured from a single stellar population model from
\citet{delg05}.  A stellar population with the age $\ga 10~\Gyr$ has
$W_0(\Hb) \approx 2~\ang$.  An \Hb\ index $\ga 3~\ang$ is expected for
a stellar population during a poststarburst phase from $\sim 10~\Myr$
up to $\sim 2~\Gyr$.
(\emph{A color version of this figure is available in the online journal.})
}
\label{fig.v_nad_vs_index_hb}
\end{figure}

The Balmer absorption lines are sensitive to the age of the underlying
stellar population and become prominent in the spectrum in which A to
early-F stars, living up to $\sim 1.5~\Gyr$, contribute significantly.
This feature has been exploited to find the signature of
poststarbursts in distant galaxies, often using \Hd\ and/or \Hc\
absorption due to less contamination from emission filling
\citep[e.g.,][]{dres83, zabl96, dres04, goto07}.  The limited spectral
baseline makes only low-lying Balmer lines available for our sample.
\fig{fig.v_nad_vs_index_hb} shows the \NaID\ velocity as a function of
the \Hb\ spectral line index $W_0(\Hb)$.  A spectral line index is an
estimate of the sum of emission and absorption equivalent widths
(\sect{Spectral Line Indices}).  In the figure, outflows are seen in
the following two classes of objects.  One class is those with $W_0(\Hb) <
0~\ang$, meaning that \Hb\ is seen in emission, so they are the
outflows seen in star-forming galaxies in the blue cloud
(\fig{fig.ub_0_vs_m_b}).  Another is those with $W_0(\Hb) \ga 3~\ang$,
slightly offset from a concentration of objects with systemic \NaID\
velocity around $W_0(\Hb) \approx 2~\ang$.  To put this in
perspective, the inset of \fig{fig.v_nad_vs_index_hb} shows the
evolution of \emph{stellar} \Hb\ absorption index over a range of
stellar ages.  A stellar population with the age $\ga 10~\Gyr$ has its
$W_*(\Hb)$ asympototing to $\approx 2~\ang$.  The \Hb\ absorption
would be seen at $W_0(\Hb) \ga 3~\ang$ during the poststarburst phase
from $10~\Myr$ to $\sim 2~\Gyr$.  In practice, a measured \Hb\ index
from a galaxy spectrum includes some nebular emission line flux
originating from the \HII\ regions surrounding hot, young stars in the
presence of residual star formation, pushing a Balmer absorption index
to a smaller, less positive value.  Therefore, $W_0(\Hb) \ga 3~\ang$
objects are likely to be in the poststarburst phase.  A large number
of systemic \NaID\ velocity objects around $W_0(\Hb) \approx 2~\ang$
is also consistent with old, quiescent galaxies not hosting outflows.

The bottom panel of \fig{fig.ub_0_and_index_hb_vs_m_nuv} shows the
relation between the $\NUV - R_{\rm AB}$ color and the \Hb\ spectral
line index for the subset of sample with both measurements.  Although
the small statistics make interpretation difficult, we can see that
the red-sequence objects with outflows appear to have a slightly
higher $W_0(\Hb)$ in general compared to those without.  A few
blue-sequence objects having $\NUV - R_{\rm AB} < 5.4$, both with and
without outflows, also move to the $W_0(\Hb) > 0~\ang$ regime.  If
these objects host residual star formation, inferred from their blue
$\NUV - R_{\rm AB}$ color, their emission fluxes may fill \Hb\
absorption, making the observed $W_0(\Hb) \approx 2$--$3~\ang$ only
lower limits to their absorption equivalent strength.  This would make
the case stronger for the poststarburst identity.

The combination of selection cuts from our sample yields a small
number of fully overlapping objects across a variety of measurements
and makes it difficult to reach generalizing conclusions with
statistical rigor.  Nevertheless we do find overwhelming consistency
in the evidence that our LIRG-type outflows are mostly seen either in
starburst or poststarburst objects.  This naturally fits into the
currently favored scenario of galaxy evolution between blue-cloud and
red-sequence galaxies, in which some mechanism (e.g., merger) triggers
a starburst in a blue galaxy which then gets ``quenched'' by some
feedback mechanism, such as by an AGN or supernovae, by the time the
galaxy joins the red sequence.  The outflows may be the result of such
feedback process observed in ``transition'' objects.

However, we must also note that stronger Balmer absorption only
indicates that there was certainly a detectable \emph{enhancement},
relative to the present, of star formation in the recent past; whether
or not an individual case makes the criteria for a conventional
\emph{starbusrt} is admittedly unclear, especially given our crude
diagnostics.  There are plausible ways for galaxies to quench star
formation without going through a starburst phase, and such mechanisms
could generate outflows and detectable enhancement of Balmer
absorption, if quenching occurs quickly.  This possibility can be
explored further only through better diagnostics.
\footnote{ {\hl With our sample, no robust way exists for directly
connecting poststarburst to the kind of star formation detected in
NUV.  The advantage of NUV diagnostics here is its sensitivity to
low-level star formation that normally eludes detection in the
visible, due to much larger light contribution from old stars to the
visible region of galaxy spectrum, and may not tell us much about the
timescale over which such low-level star formation has been in
existence.  Therefore the star formation seen in NUV could be
``residual'' star formation from a recent starburst/star-forming event
or it could just be a small amount of continuous star formation.}  }

\subsection{Host Morphology}
\label{Host Morphology}

\begin{figure}
\scalefigure
\plotone{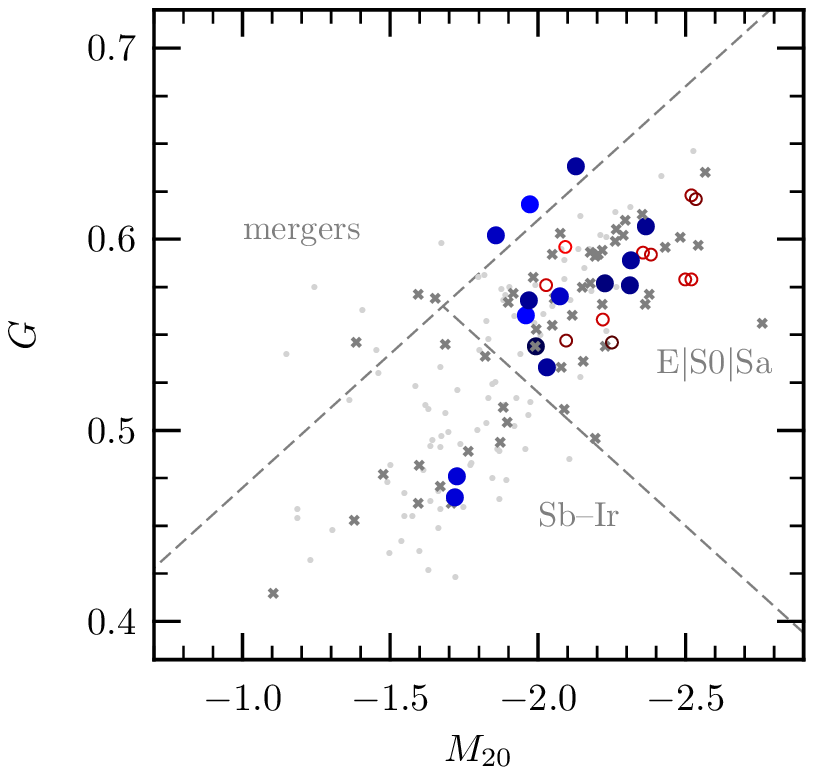}
\caption{ Quantitative morphology measure (Gini coefficient $G$ and
the second-order moment of the brightest $20\%$ of the total galaxy
flux $M_{20}$) from the analysis of \emph{HST}/ACS images by
\citet{lotz06a}.  The morphological cuts (\emph{dashed gray lines}) to
the ``traditional'' classes are from \citet{lotz06a}, calibrated with
the $z = 0$ rest-frame $B$-band morphology of their sample.  The
symbols are as in \fig{fig.snr_index_nad_vs_csnr}.
(\emph{A color version of this figure is available in the online journal.})
}
\label{fig.gini_vs_m20_w_kinematics}
\end{figure}

\begin{figure}
\scalefigure
\plotone{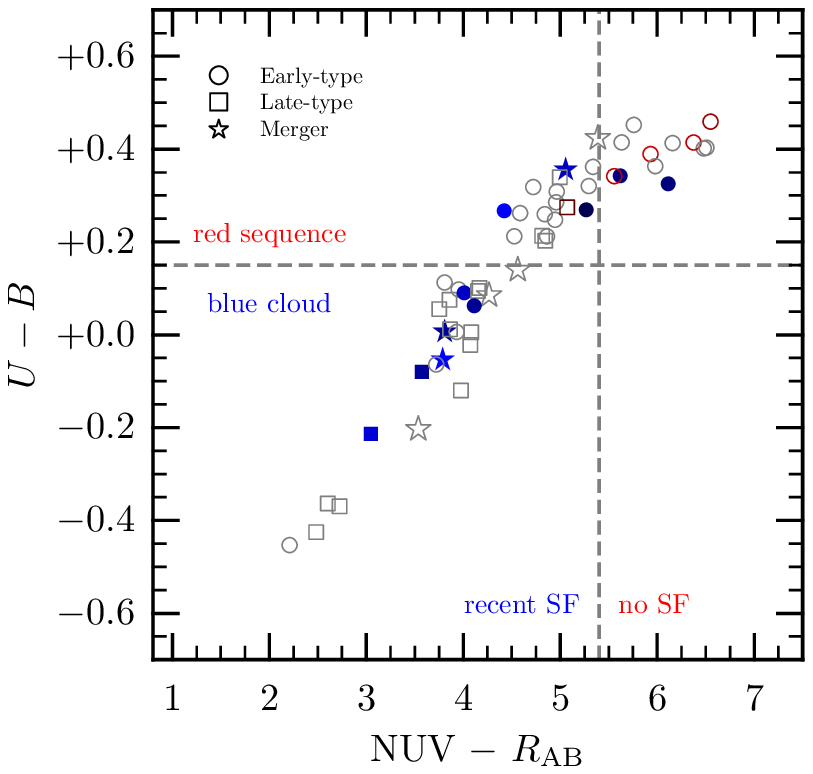}
\caption{ Morphology of objects in terms of the rest-frame $(U-B,
\NUV-R_{\rm AB})$ color-color diagram; see
\fig{fig.ub_0_and_index_hb_vs_m_nuv}.  The morphology, from the
analysis of \emph{HST}/ACS images by \citet{lotz06a}, is divided into
early-type (E/S0/Sa; \emph{circle}), late-type (Sb--Ir;
\emph{square}), and merger-candidates (\emph{star}).  The filled symbols
indicate host galaxies of outflows.  The symbol colors are as in
\fig{fig.snr_index_nad_vs_csnr}.  Only the objects from the high-\snr\
\NaID\ velocity sample (\sect{NaID Velocity and Blueshift
Probability}) with \NUV\ measurements are plotted.
(\emph{A color version of this figure is available in the online journal.})
}
\label{fig.ub_0_vs_m_nuv_w_qmorph}
\end{figure}

\fig{fig.gini_vs_m20_w_kinematics} shows a subset of galaxies with
\NaID\ measurements for which we have the quantitative morphology from
the \emph{HST}/ACS imaging analysis by \citet{lotz06a}.  A set of new
nonparametric morphology measures, Gini coefficient $G$ which measures
the distribution of flux among pixels over a galaxy and the
second-order moment of the brightest $20\%$ of the total galaxy flux
$M_{20}$, is calibrated visually to the traditional morphological
classifications \citep{lotz04}: early-type (E/S0/Sa), late-type
(Sb--Ir), and merger candidates.  Our survey is more sensitive to
luminous, high surface brightness galaxies (\sect{Sample Selection}),
and \fig{fig.gini_vs_m20_w_kinematics} shows that the sampling is
biased against objects with late-type (i.e., lower surface brightness)
morphology; see \citet{lotz06a} for the analysis of complete samples
drawn from the parent AEGIS survey, which shows that the many objects
with low surface brightness, late-type morphology, which would appear
toward the bottom-left corner, did not make our selection cut.

The transition between a LIRG to a ULIRG is physically plausible
in the merger sequence \citep{sand04}, so the high outflow detection
rate ($3/6 = 0.5 \pm 0.2$) in merger candidates is not surprising.
Outflows are detected in almost all local ULIRGs \citep{mart05,
mart06, rupk05p2}.  Although all three merger candidates with outflows
are detected at $24~\mu{\rm m}$, only one is a LIRG.  Since the
dynamical timescale of merger, $\sim 1~ \Gyr$, exceeds that of gas
consumption, $\sim 100~\Myr$, it is likely that we miss the (U)LIRG
phase due to its short duty cycle.  It should be noted, however, the
nature of the association between mergers/interactions and (U)LIRGs is
not as well established for objects at $z \sim 0.5$.


A majority ($9/14 = 0.64 \pm 0.13$) of outflows are seen in the
objects with early-type morphology (E/S0/Sa).  An insight on the order
of morphological transformation may be given by
\fig{fig.ub_0_vs_m_nuv_w_qmorph}, which plots the distribution of
morphology in terms of the rest-frame $(U-B, \NUV-R_{\rm AB})$
color-color diagram.  Consistent with the common knowledge, most
late-type galaxies are in the blue cloud, while most early-type
galaxies are in the red sequence.  The distribution of outflow hosts
roughly follows that of the parent sample.  Although the small sample
size does not allow a statistically robust conclusion, outflows appear
to be seen more preferentially where merger candidates are also seen.
Previously we find evidence that some outflows are seen in
poststarburst objects (\sect{Evidence for Poststarburst}).  The
early-type morphology, coupled with the signs of interaction in a few
of the red-sequence outflows, is consistent with the existing studies
of poststarburst galaxies as \emph{transition} objects in the
morphology sequence, which conclude that they are spheroidal, often
showing signs of interaction \citep[e.g.,][]{yang04, goto05}.

We find only two outflows with late-type morphology, at $z > 0.35$.
\fig{fig.gini_vs_m20_w_kinematics} makes clear that our selection is
highly biased toward high surface brightness galaxies; we have missed
a large population of low surface brightness star-forming galaxies
that are dim in the rest-frame visible continuum, with which our
primary selection cut was made.  These outflows with late-type
morphology are detected significantly in the infrared, having
$\log{(\Lir / L_\odot)} \approx 11.1$ and $10.4$.  Unlike ULIRGs, the
morphology of LIRGs appears to be a mixed bag.  At $z \sim 0.7$,
however, more than half of their LIRGs have disk morphology
\citep[e.g.,][]{bell05, melb05}, suggesting that the abundant LIRG
population simply reflects an elevated level of star formation in
normal galaxies at an earlier lookback time.  All our objects are at
lower redshifts, and it is unclear if we are observing disky LIRGs of
the type seen at higher redshifts.

\section{Discussion}
\label{Discussion}

We have seen that the detection rate of LIRG-like outflows is a strong
function of infrared luminosity or star formation rate (\sect{Trends
with Star Formation}).  The distribution of specific star formation
rates at a fixed stellar mass (\fig{fig.ssfr_vs_mass_star}) might
imply that the frequency of outflows may be higher for the objects in
which ongoing star-forming activity is enhanced relative to the past
average (i.e., high birthrate parameter).  The distribution in the
optical color magnitude diagram shows that outflows are abundant both
in the blue cloud and the red sequence (\sect{Color Magnitude
Diagram}).  The NUV photometry reveals that the red-sequence outflows
are hosted by the objects that have gone through recent star formation
(\fig{fig.ub_0_and_index_hb_vs_m_nuv}).  There is evidence that some
of the red-sequence outflows are in dusty (\fig{fig.ub_0_vs_beta}),
star-forming galaxies (\fig{fig.ub_0_vs_m_b_w_lir}).  Yet some may be
poststarburst galaxies (\sect{Evidence for Poststarburst}) just
arriving the red sequence, which is consistent with their early-type
morphology (\sect{Host Morphology}).  We now discuss a few fine points
that need attention, and describe how our finding of \NaID\ outflows
might fit into the current understanding of galaxy formation and
evolution.

\subsection{Redshift Evolution of Outflows}
\label{Redshift Evolution of Outflows}

\fig{figs.mass_star_vs_z} shows the detection of outflows as a
function of redshift in our sample.  Over the redshift range,
star-forming properties of galaxies are expected to change and so do
the detection rates of outflows.  In the figure, no $M_* \approx
10^{11} M_\odot$ galaxies at $z \approx 0.2$ show signs of an outflow,
yet a significant fraction of galaxies with the similar mass at $z
\approx 0.5$ do.  This is consistent with the known trend of the star
formation history of galaxies and the increasing trend of outflow
detection rate with SFR (\sect{Trends with Star Formation}).
\citet{noes07a} shows that star-forming galaxies with $M_* \approx
10^{11} M_\odot$ are rare at $z < 0.45$ but are abundant and most are
LIRGs at $z > 0.45$.  \fig{figs.mass_star_vs_z} shows that the
outflows detected in our sample are typically seen in high-$M_*$
galaxies at higher $z$ with an SFR expected in a LIRG ($\ga
20~M_*~\yr^{-1}$).  Due to low \snr, however, we cannot say whether or
not lower-$M_*$ galaxies at $z \sim 0.5$ host outflows.  Nevertheless,
at the high $M_*$ end, we clearly observe that the ``downsizing''
effect extends to the detection rate of LIRG-like outflows, i.e., the
typical mass of galaxies that host outflows move to a lower mass
toward lower redshift, much like the trend seen in cosmic star
formation history of galaxies \citep[e.g.,][]{mada96, lill96, lefl05,
fabe07, noes07a, coop07b}.  \citet{bund06} qunatify the evolving
trend in terms of the ``quenching mass limit,'' which is shown in
\fig{figs.mass_star_vs_z} for a crude comparison.  An important
connection is likely to exist between quenching events and gaseous
feedback, so the redshift evolution of outflows needs to be explored
in view of host galaxy properties.

\subsection{Merger-Triggered Activities and Outflows}

Merger-driven galaxy evolution has been a prominent paradigm over the
past decades, and the current success of the \LCDM\ theory in
describing the hierarchical structure formation certainly points to
its significant roles in galaxy-scale phenomena.  The relevance of
mergers in shaping the cosmic star formation history of galaxies,
however, is under intense scrutiny.  The difficulty arises from the
fact that direct identification of mergers and quantifying their
frequency from the observation of high-redshift galaxies remains very
challenging, due to a number of factors including surface brightness
dimming and shifting passbands.  Galaxy mergers of star-forming disk
galaxies have attracted much attention, because they are a mechanism
widely known to cause enhanced star formation, after which merger
remnants dynamically relax into spheroids \citep[e.g.,][]{miho96,
cox06}.  Along with the expected increase of the merger rate with the
lookback time from numerical simulations \citep[e.g.,][]{gott01},
mergers have been invoked as a mechanism responsible for the declining
trend of the comoving star formation density since $z \sim 2$
\citep[e.g.,][]{lill96, mada96}, although exactly how much mergers
contribute to the trend is still debated \citep[e.g.,][]{brid07,
lotz06a}.  Galaxy merger is appealing also for it affects star-forming
as well as morphological properties of galaxies in ways that naturally
explain the observed transition of young (i.e., blue and disky) into
old (i.e., red and spheroidal) galaxy populations.  In principle, a
variety of such transient phenomena as starburst and quasars can also
be integrated into a coherent picture of merger-driven galaxy
evolution \citep[e.g.,][]{sand96, hopk06}.

The correlation of the outflow detection rate with the degree of
elevation in star-forming activity certainly indicates that starbursts
are an important component of galactic superwind phenomenon.  However,
outflow is exciting not because it is associated with such a spectacular
starburst event, but because it may carry away from the host galaxy
the bulk of ``fuel'' for further star formation, which may contribute
to the ``quenching'' of star formation.  In fact, ULIRGs (i.e.,
gas-rich mergers) show dynamical evidence for spheroids in formation
\citep[e.g.,][]{genz01}, and the star formation history of
poststarburst galaxy is consistent with the (U)LIRG origin
\citep[e.g.,][]{pogg00, bekk01, kavi07}.  Our detection of \NaID\
outflows in poststarburst galaxies strongly suggests that feedback
mechanism affects the kinematics of cool interstellar gas well after
the most intense phase of star formation.  The fact that quite a few
red-sequence galaxies host winds may not be surprising yet still a
striking result.  A vast majority of absorption-line studies of
outflows in literature has been on vigorously star-forming systems.
Using a plot similar to \fig{fig.index_mgb_vs_index_nad},
\citet{rupk05p1} showed that most outflows in infrared-selected
galaxies were detected in the loci of low \MgIb\ and high \NaID\
absorption indices, i.e., young galaxies with high interstellar \NaID\
column (their Fig.~8).  In contrast, the red-sequence galaxies with
outflows in our sample can have a high \MgIb\ absorption index,
reflecting the presence of intermediate to old stellar population, and
are not clearly distinct from other red-sequence galaxies in an
optical color magnitude diagram (\fig{fig.ub_0_vs_m_b}).  The apparent
connection between outflows and high $M_*$ poststarburst galaxies
provides circumstantial evidence that outflows of cool gas contributes
to or is a consequence of more effective quenching of star formation.

How the progenitors of spheroidal galaxies in the local universe
evolve into their current state remains a topic under vigorous
investigation.  While their stellar contents suggest that massive
spheroids have been passively evolving and that their stellar mass
changes little since $z \sim 1$ \citep[e.g.,][]{brin00, bund05}, the
evolution of the luminosity function suggests that the number density
of luminous red galaxies has increased by a factor of $\sim 2$ over
the same period \citep[e.g.,][]{bell04, brow07, fabe07}.  Over the
similar redshift range, star-forming galaxies have a relatively small
spread in SFRs at a fixed mass in the blue sequence
\citep[e.g.,][]{noes07a}, leading to a paucity of objects presumably
in transition between the blue and red sequences.  Furthermore, less
massive spheroids have younger stellar contents
\citep[e.g.,][]{treu05, kavi08} and the characteristic galaxy mass
above which the star formation in galaxies quenched evolves over
redshift in a downsizing fashion \citep{bund06}, indicating that
catastrophic transition events occur at a progressively lower mass
toward lower redshift.  The exact rate of transition is very difficult
to estimate, yet indirect arguments favor rare and/or fast
mechanism(s) \citep[e.g.,][]{blan06}.  Gas-rich mergers perhaps
contribute to some but not all of these blue-red transitions
\citep{bund07}.  The qualitatively similar downsizing trend in
star-forming galaxies and outflow hosts over $z < 0.5$ (\sect{Redshift
Evolution of Outflows}), however, may imply that the mechanism
responsible for downsizing of star formation may also accompany
outflows.  A rigorous conclusion must await the quantitative analysis
of the sample which suffers less from small number statistics,
selection effects, and incompleteness.

We must also note that it is not clear that merger per se is a
necessary precursor for outflows.  The direct morphological
evidence for interaction is not very strong in our outflows (a
majority of outflows are in early-type galaxies; \sect{Host
Morphology}).  A circumstantial evidence is provided only through an
indirect argument that a spheroidal formation follows a merger-induced
starburst with a poststarburst signature.  Since the timescale for
mergers at high redshift to remain identifiable is shorter than a
poststarburst phase, the scarcity of direct evidence may not
immediately discount the importance of mergers.  However, it is
possible that low-level star formation in early-type galaxies could
drive outflows, perhaps via the accretion of gas-rich satellites or
minor mergers.  Early-type galaxies also have reservoirs of hot gas,
which could provide fuel for star formation via condensation.  It has
been suggested that the early-type galaxy population itself shows
bimodality in their UV-visible color distribution, reflecting the
richness in their star formation histories; low-level star formation
appears to continue in less-massive early-type galaxies
\citep{kavi08}.  The origin of outflows in the red galaxies may not be
as simple as a quenching event followed by passive evolution.

Furthermore, the mounting evidence now shows that a majority of
$z \sim 1$ LIRGs are normal disk galaxies whose elevated star
formation is \emph{not} caused by interaction \citep[e.g.,][]{bell05,
lotz06a}.  The gradual decline of star-forming efficiency in their
disks may be responsible for much of the global trend seen in comoving
SFR density.  From our study alone, whether or not these disky LIRGs
at high-redshift host outflows is not clear; the nature of
outflows in these objects may be different from merger-induced ones.

It would be interesting to see how the presence of outflows at $z \la
1$ contribute to the fate of $z \sim 1$ LIRGs down to $z \sim 0$,
which may either stay but fade gradually within the blue sequence or
go through rapid quenching of star formation to migrate to the red
sequence.  Given the strong evolution of galaxy properties in general
(\sect{Redshift Evolution of Outflows}), our knowledge from the local
study of LIRG-like outflows might not be relevant for high-$z$, disky
LIRGs.  On the other hand, if the outflow strength correlates with
some parameter such as the presence of nuclear activity or their
degree of interaction in relation to their morphology among high-$z$
LIRGs [as observed in local ULIRGs by \citet{mart05}], deprivation of
(cool) gas via superwind may be important in transforming star-forming
disks into quiescent spheroids in catastrophic events at those
redshifts.  The detection rate of outflows in view of host galaxy
morphology at $z \la 1$ may provide some insight on the physical
mechanism that maintains the bimodality in galaxy population since $z
\sim 1$.  A comparison of mass outflow rates in $z \sim 1$ LIRGs with
different morphology would also make an interesting exercise to put
some constraint on the role of mass loading in the evolutionary
histories of morphology and star formation.  Fortunately, several
useful UV resonance lines at different ionization states shift into
visible window for $z \sim 1$ objects \citep[e.g.,][]{wein08},
so future surveys are in a better position to constrain mass loading
from these lines \citep[e.g.,][]{murr07}.

\subsection{Star Formation Versus AGN}
\label{Star Formation versus AGN}

It is usually assumed that superwinds are driven by thermalized energy
output from supernovae.  In principle, AGNs offer much larger
repository of energy than supernovae, and the ubiquity of
supermassive black holes in galactic spheroids inferred from the
$M_{\rm bh}$-$\sigma_v$ relation \citep[e.g.,][]{trem02} and their
co-evolution with the lookback time \citep[e.g.,][]{woo06} suggests
that the energy output from AGNs may play important roles in galaxy
formation.  While AGNs offer increasingly attractive solutions to the
yet elusive mechanism for shutting off star formation in massive
objects, where and how their energy output couples to the gas in and
around galaxies remains to be identified.  Recently, however, several
observational studies have elucidated the connection between
poststarburst and nuclear activity: host galaxies of quasars often
show poststarburst signature in their continuum emission
\citep[e.g.,][]{cana06}; the morphology of poststarburst galaxies
often show a blue core, which produces a LINER spectrum
\citep{yang06}; the optical emission-line ratios classify a majority
of poststarburst galaxies into Seyfert/LINERs \citep{trem05, yan06,
yan06aas}.

\begin{figure}
\scalefigure
\plotone{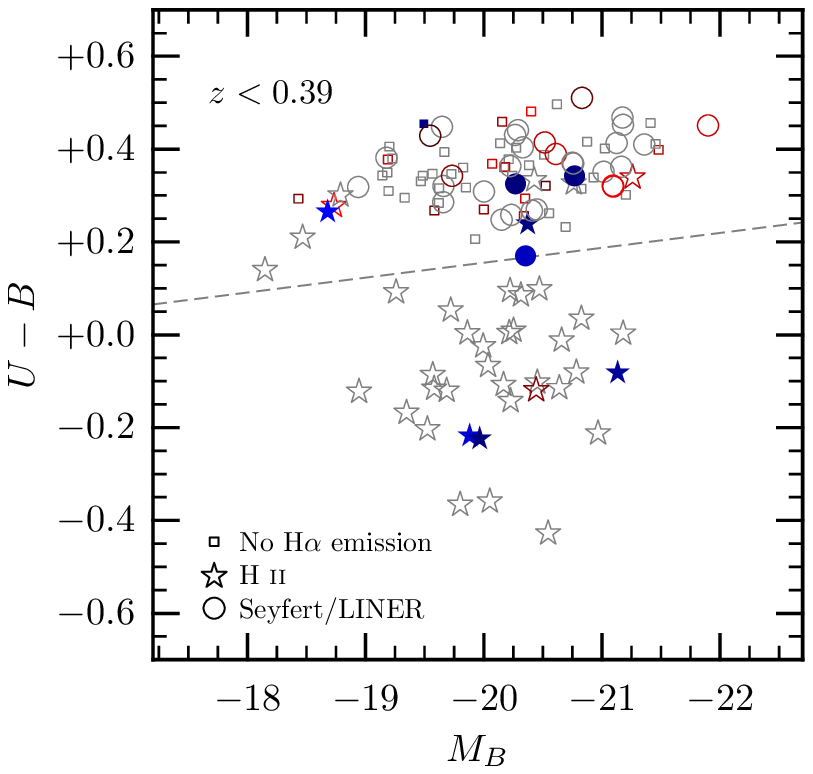}
\caption{ Rest-frame ($U-B$, $M_B$) color-magnitude diagram for the
objects in the high-\snr\ sample with $\fNII/\Ha$ emission equivalent
width ratio; see \sect{Star Formation versus AGN}.  The spectral
baseline restricts the $\fNII/\Ha$ measurements to $z < 0.39$ objects.
The symbols indicate the equivalent width ratio of Seyfert/LINER
[$\log{(\fNII/\Ha)} \ge -0.3$; \emph{circle}] and \HII-region
[$\log{(\fNII/\Ha)} < -0.3$; \emph{star}]; the objects without
detectable \Ha\ emission do not have equivalent width ratios
(\emph{square}).  The colors are as in
\fig{fig.snr_index_nad_vs_csnr}, and outflows are marked by filled
symbols.
(\emph{A color version of this figure is available in the online journal.})
}
\label{fig.ub_0_vs_m_b_w_ha_n2}
\end{figure}

Low-ionization outflows are found to be faster in starburst galaxies
with some indication of AGN \citep{mart05, rupk05p2}.  In $z \sim 0.5$
poststarburst galaxy sample, \citet{trem07} find \MgII\ outflows of
$\sim 1000~\km~\s^{-1}$, much faster than the typical \NaID\ outflows
in ULIRGs ($\sim 300$--$400~\km~\s^{-1}$).  These fast outflows in the
galaxies at their \emph{postquasar} phase provide compelling evidence
for AGN-driven outflows in poststarbursts.  It is worth noting that,
due to the difference in sample selection, our poststarburst outflows
in general are likely to be of a more typical kind than those of
\citeauthor{trem07}, i.e., the event which led to the suppression of
star formation does not have to accompany quasar activity for us to
classify them as poststarbursts.  Although the exact values are
physically meaningless (\sect{Definition of Outflow}), our
measurements imply \NaID\ outflow velocities of $\sim
100~\km~\s^{-1}$, which do not favor one scenario over others as to
the outflow driving mechanism.  It is plausible that some of our
red-sequence outflows have gone through the kind of postquasar phase
that \citeauthor{trem07} observed, but they do not all have to be.

Nonetheless, the fact that typical poststarburst galaxies show
low-level nuclear activity seems convincing, given the recent studies
of the line emission from red-sequence galaxies.  In
\fig{fig.ub_0_vs_m_b_w_ha_n2}, we show the classification of
Seyfert/LINER versus star formation using the $\fNII/\Ha$ emission
\emph{equivalent width} ratio, such that $\log{(\fNII/\Ha)} = -0.3$
divides the two classes.\footnote{Unfluxed DEEP2 spectra force us to
use equivalent widths in place of line fluxes, yet the adjacent lines
forming the ratios are so close that they give similar results
\citep{kobu03}.  The \Ha-\fNII\ complex is fitted with three Gaussians
for emission lines.  When \Ha\ absorption is obviously present,
another Gaussian is fitted; otherwise, a fiducial rest-frame
equivalent width of $2 \pm 1~\ang$ is added to the \Ha\ emission
flux.} The spectral baseline restricts the $\fNII/\Ha$ measurements to
$z < 0.39$ objects.  The use of $\fOIII/\Hb$ allows us to extend the
classification to $z > 0.39$ objects in which we see a majority of
outflows, but the delineation of Seyfert/LINER and star formation
becomes notoriously ambiguous when only that ratio is used.  A
comparison to \citet{wein07} shows that \emph{all} $z > 0.35$ outflows
in our sample have \HII-region--like ratios in $\fOIII/\Hb$, which is
rather puzzling, given many of them are red galaxies where line
emissions tend to originate from AGN/LINERs, according to local
studies.  Using $\fNII/\Ha$, \fig{fig.ub_0_vs_m_b_w_ha_n2} shows a
distribution of objects grossly consistent with \citet{yan06}; i.e.,
line emission from the red sequence is dominated by Seyfert/LINER,
while that from the blue cloud is mostly from star formation.  Host
galaxies of outflows appear to follow the parent distribution, yet the
analysis suffers from small number statistics.

\begin{figure}
\scalefigure
\plotone{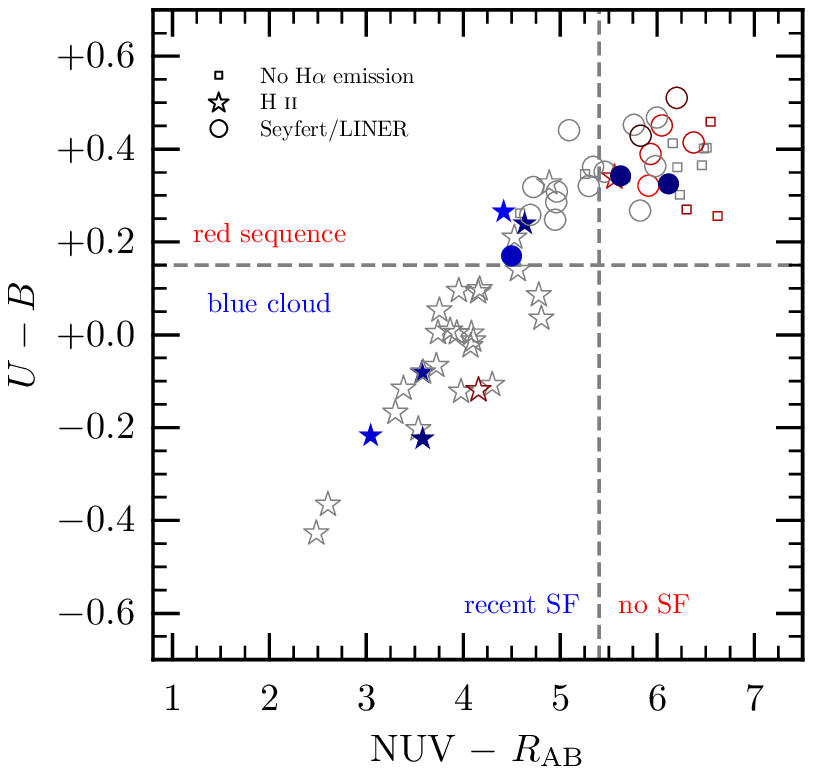}
\caption{ Rest-frame $(U-B, \NUV-R_{\rm AB})$ color-color diagram
for the objects in the high-\snr\ sample with $\fNII/\Ha$ emission
equivalent width ratio.  Symbol shapes are as in
\fig{fig.ub_0_vs_m_b_w_ha_n2}.  Symbol colors are as in
\fig{fig.snr_index_nad_vs_csnr}, and outflows are marked by the filled
symbols.  The photometric color delineations are described in the
caption of \fig{fig.ub_0_and_index_hb_vs_m_nuv}.  The spectral
baseline restricts the $\fNII/\Ha$ measurements to $z < 0.39$ objects.
(\emph{A color version of this figure is available in the online journal.})
}
\label{fig.ub_0_vs_m_nuv_w_ha_n2}
\end{figure}

In \fig{fig.ub_0_vs_m_nuv_w_ha_n2}, we have a slightly different view
on the color distribution of galaxies in terms of the emission
excitation.  The advantage of adding the near-UV photometry to detect
low-level star formation has been discussed in \sect{Color Magnitude
Diagram}.  The small number of outflow host galaxies at $z < 0.39$
still limits our interpretation.  Nevertheless, it is quite
interesting that the transition from star formation to Seyfert/LINER,
and then to no \Ha\ emission appears to happen along the stellar age
sequence, and the abundance of Seyferts/LINERs is observed in the
region occupied by the galaxies in transition, a significant fraction
of which appears to host outflows at $z \sim 0.5$
(\fig{fig.ub_0_and_index_hb_vs_m_nuv}).  Although the temporal
co-existence of relic outflows and current nuclear activity does not
necessarily imply any causal connection, extending this analysis to
higher redshift, where more outflows are expected, could help us
understand the potential impact of winds in establishing the observed
trend in terms of their driving mechanisms.

The presence of inflows in red, early-type
(\fig{fig.gini_vs_m20_w_kinematics}) galaxies in the above figures is
also remarkably consistent with their host either being quiescent or
having their optical emission lines excited by AGN/LINERs, with little
evidence for star formation.  It remains to be seen whether any
evidence exists to directly connect the inflows and the nuclear
activities.  On the other side of outflows, the nature of inflows also
needs to be explored, since it has been long speculated that some
mechanism, an AGN being a prime suspect, is preventing more stars from
being formed than observed in massive elliptical galaxies.

\section{Summary}
\label{Summary}

We reported on a \snr-limited search for low-ionization
outflows using the DEEP2 spectra of the $0.11 < z < 0.54$ objects in
the AEGIS survey.  Doppler shifts from the host galaxy redshifts were
systematically searched for in the \NaID\ optical resonance absorption
doublet.  This was the very first time that a signature of galactic
superwind was systematically looked for in the individual galaxy
spectra from a modern, large spectroscopic redshift survey.

Our \NaID\ profile fitting method closely followed that of
\citet{rupk05p1}, explicitly fitting the absorption line model
parameterized by the wavelength and the optical depth at the line
center, Doppler width, and covering fraction in a self-consistent
manner.  The confidence intervals in the \NaID\ velocities were
estimated through the MCMC sampling technique.  This allowed us to
evaluate the quality of \NaID\ velocity measurements visually in terms
of probability distributions of model parameters.  Although the
spectral resolution and \snr\ limited us to studying the
interstellar gas kinematics by fitting a single doublet component to
each observed \NaID\ profile, LIRG-like outflows should have been
detected at $\ga 6\sigma$ in absorption equivalent width down to the
survey limiting \snr\ ($\sim 5$ pixel$^{-1}$) in the
continuum around \NaID.  This meant that, if a \NaID\ outflow of a
comparable strength to the LIRGs detected by the \citet{rupk05p2}
survey left its signature in a spectrum, we were able to detect the
presence in our survey.  We discussed the challenges involved in
recovering physically important parameters from a moderate \snr\
spectrum yet argued that a Doppler shift could be measured robustly.

The detection rate of LIRG-like outflow clearly showed an increasing
trend with star-forming activity and infrared luminosity.  However, by
virtue of not selecting our sample on star formation, we also found a
significant fraction of outflows in galaxies on the red sequence in
the rest-frame ($U-B$, $M_B$) color-magnitude diagram.  Most of these
red-sequence outflows were of early-type morphology and showed the
sign of recent star formation in their UV-optical colors; some showed
enhanced Balmer \Hb\ absorption lines indicative of poststarburst as
well as high dust extinction.

We also note that inflows were detected in some red, early-type
galaxies.  Although the definition of an inflow in this study was just
symmetrically opposite to that of an outflow, the fact that we
observed them in very distinct subsets of galaxies strongly indicated
that our outflow/inflow detections were not spurious.  However,
with the difficulty in removing stellar absorption component at \NaID\
as well as the lack of immediate explanations for their driving
mechanisms, the investigation of inflows was beyond the scope of this
paper and will be explored in future AEGIS studies.  We merely note
that a connection is suspected between an inflow and the feeding of
nuclear activity in the maintenance mode of AGN feedback.

The fact that many of our outflows have been found in galaxies
presumably in transition suggests that galactic superwinds could
outlive starbursts and play a role in quenching star formation in the
host galaxies on their way to the red sequence.  The detectable
imprints of gaseous feedback in these galaxies provide us a means to
observationally constrain different feedback models.  Despite that
the small number statistics as well as selection effects hindered our
ability to rigorously characterize the nature of the host galaxies of
outflows across a wide array of physical parameters accessible in the
AEGIS survey, the initial analysis presented in this paper will help
design future experiments on the impact of superwinds on galaxy
evolution in the epoch when the star-forming properties of galaxies
drastically change since $z \sim 1$.  Gaseous kinematics adds to
future studies a useful dimension to explore and opens up a promising
avenue for constraining the driving mechanisms of baryons cycling
through the components that constitute the luminous part of the
universe, as well as for directly quantifying how much gas joins in
such process.

\acknowledgments

T.S.~would like to thank the following scientists for inspirations:
Christy Tremonti for sharing her stimulating results on SDSS \NaID\
outflows, as well as her hospitality during his visits to Steward;
Philip Marshall for enlightenment with the Bayesian statistical
methods; and David Rupke and his coworkers for the series of detailed
work on \NaID\ outflows.  T.S. gratefully acknowledges Alison Coil for
her thorough reading of the manuscript and her very insightful
suggestions and also wishes to extend his thanks to Ben Weiner, Sandy
Faber, and Fran\c{c}ois Schweizer for helpful discussions.  We thank
the anonymous referee for his or her thorough reading, constructive
feedback, as well as patience.  This work was never possible without
the dedicated efforts, contributions, as well as the generosity from
all the AEGIS/DEEP2 members.

The research presented in this paper made an extensive use of the
Python programming language and the associated tools.  PyRAF and
PyFITS are products of the Space Telescope Science Institute, which is
operated by AURA for NASA.  The figures in this paper are all prepared
by PyTioga\footnote{PyTioga is available at
http://pytioga.sourceforge.net/.}, an open source software for
creating figures and plots using Python, PDF, and TeX.

This research has made use of the NASA Astrophysics Data System
abstract service.

Financial support was provided by the David and Lucille Packard
Foundation.

Funding for the DEEP2 survey has been provided by NSF grant
AST-0071048 and AST-0071198.  Some of the data presented herein were
obtained at the W.M. Keck Observatory, which is operated as a
scientific partnership among the California Institute of Technology,
the University of California and the National Aeronautics and Space
Administration.  The Observatory was made possible by the generous
financial support of the W.M. Keck Foundation.

We gratefully acknowledge NASA's support for construction, operation,
and science analysis of the \emph{GALEX} mission, developed in
cooperation with the Centre National d'Etudes Spatiales of France and
the Korean Ministry of Science and Technology.

For the full acknowledgement of the AEGIS data set, please refer to
\citet{davi07}.

Last but not least, we recognize and acknowledge the very
significant cultural role and reverence that the summit of Mauna Kea
has always had within the indigenous Hawaiian community.  We are most
fortunate to have the opportunity to conduct observations from this
mountain.


\appendix
\section{Modeling the \NaID\ Absorption Line}

We closely follow the model presented by \citet{rupk02, rupk05p1};
readers are strongly encouraged to refer to these papers for a
thorough discussion.  Only a single component of the \NaID\ doublet is
fitted in each spectrum; that is, we assume that our fitting of one
absorption doublet is sensitive to the bulk property of the multiple
\NaID\ absorbing clouds integrated along the sightline.  While this is
certainly an oversimplification \citep[see, e.g.,][for how complex an
absorption profile with several kinematic components can
appear]{rupk02}, the limited \snr\ does not allow more detailed
analysis; for the cases in which observed \NaID\ absorption lines are
well defined, the fitting results are reliable.  For the sole purpose
of detecting a Doppler shift from a systemic redshift (and \emph{not}
measuring the exact velocity value), a single component fit is an
acceptable compromise.  We are effectively addressing whether an
observed \NaID\ line profile can accommodate a single Doppler-shifted
\NaID\ absorber.

The observed intensity profile $I(\lambda)$ of an absorption line is
fitted by a model profile parameterized in the optical depth space.
That is, if the (velocity-independent) partial covering fraction is
\Cf, each \NaID\ doublet is modeled by
\[
I(\lambda) = 1 - \Cf \left[1 - e^{-\tau_{\rm blue}(\lambda)-\tau_{\rm
red}(\lambda)} \right] \ ,
\]
where $\tau_{\rm blue}(\lambda)$ and $\tau_{\rm red}(\lambda)$ are the
optical depths of blue and red components of the doublet as a function
of wavelength $\lambda$.  (This expression is appropriate for a
continuum-divided spectrum.)  We assume that the velocity distribution
of absorbing atoms within a cloud is Maxwellian, such that each
absorption line is modeled as
\[
\tau(\lambda) = \tau_0 e^{-(\lambda-\lambda_0)^2/(\lambda_0b_D/c)^2}
\ ,
\]
where $\tau_0$ is the optical depth at the line center $\lambda_0$,
$c$ the speed of light, and $b_D$ is the Doppler parameter in units of
speed.  Since the ratio of oscillator strengths for blue and red sides
of the \NaID\ doublet is two \citep{mort91}, we may assume that the
central optical depths are related via $\tau_{0,{\rm blue}} / 2 =
\tau_{0,{\rm red}}$.  Hence the intensity profile of each \NaID\
doublet component is
\[
I(\lambda) = 1 - \Cf
\left\{ 1 - \exp{\left[
-2\tau_0 e^{-(\lambda -\lambda_{\rm blue})^2
/(\lambda_{\rm blue} b_D / c)^2}
- \tau_0 e^{-(\lambda - \lambda_{\rm red})^2
/(\lambda_{\rm red} b_D / c)^2}
\right]} \right\} \ ,
\]
where $\lambda_{\rm blue} = 5889.9512~\ang$ and $\lambda_{\rm red} =
5895.9243~\ang$ are the rest-frame central wavelengths (in air) of
blue and red lines of \NaID\ doublet.  As noted by \citet{rupk05p1},
the Maxwellian velocity distribution and velocity-independence of
partial covering fraction are significant assumptions.  In case there
is an outflow, an observed \NaID\ absorption profile likely arises
from several absorbing clouds entrained in a superwind
(e.g., A. Fujita et al. 2008, in preparation), so their bulk kinematics would not be
described simply by a Maxwellian distribution.  Furthermore, if for
example a spherical virialized cloud cuts through a sightline, the
covering fraction must depend on the velocity of the constituent
particles in the cloud; thus the assumption of velocity independence
for covering fraction is not physically consistent in detail.
Nevertheless, our analysis does not benefit from relaxing these
constraints, as the quality of data does not warrant that such
detailed information can be extracted.

Before the above model is fitted, each spectrum around \NaID\ is
divided by the pseudocontinuum, a straight line fitted to the
variance-weighted spectra within the rest-frame regions of
$5822$--$5842~\ang$ and $5910$--$5930~\ang$; the ranges are
chosen after visual inspection for the best continuum normalization
centered around \NaID, while avoiding other prominent stellar
absorption features as much as possible.  In a very strong outflow,
the nebular emission line $\HeI~\lam~5876$ can contaminate the bluest
wing of a Doppler-shifted \NaID\ component.  Visual inspection
indicates that few spectra suffer from such a contamination, so no
account is taken for the \HeI\ emission in our measurements.

We take $\lambda_{\rm red}$, $b_D$, $\tau_0$, and $\Cf$ as model
parameters to be estimated via the Metropolis-Hastings algorithm
\citep{metr53, hast70}, which yields a more robust confidence interval
on a model parameter than that derived from a covariance matrix, when
the probability distribution of the model parameter cannot be
described as Gaussian.  The sampling method also improves over
\citeauthor{rupk05p1} in a sense that each Monte Carlo realization is
not drawn from the ``best'' model which needs to be chosen a priori
via least-square fitting, for example.  The priors for the model
parameters are assumed to have a uniform probability density over
their upper and lower limits: $42~\km~\s^{-1} < b_D <
700~\km~\s^{-1}$, $0 < \tau_0 < 10^3$, $0 < \Cf < 1$, and $\lam_0$ is
constrained to be within $\pm 700~\km~\s^{-1}$ of the systemic
velocity.  The \chisq\ probability distribution for the given degrees
of freedom (i.e., the number of data points fitted minus the number of
model parameters) is assumed for the likelihood function.  Each
sampling consists of $10^5$ iterations.

\begin{figure}
\scalefigure
\plotone{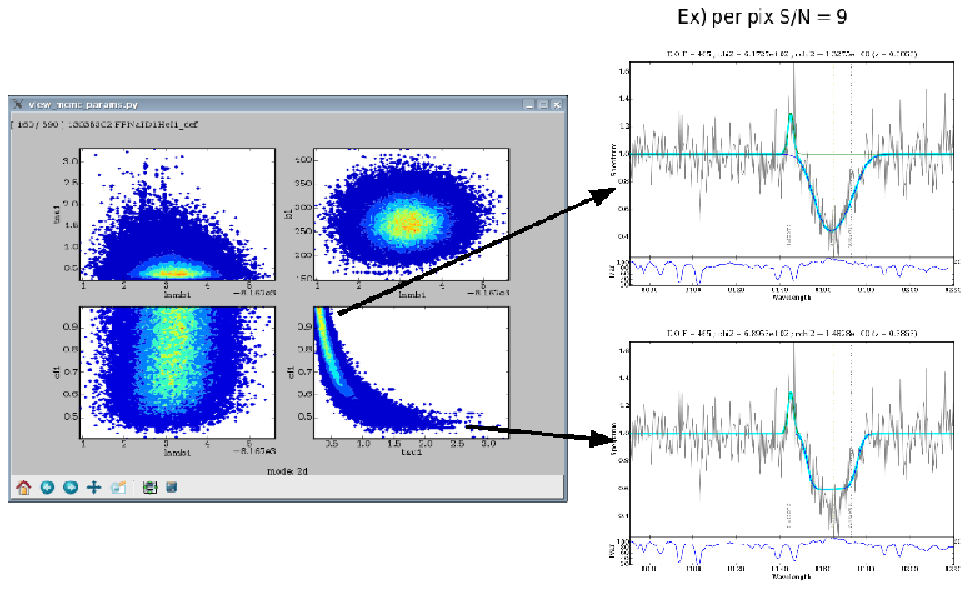}
\caption{ Few output windows from the \NaID\ measurement pipeline.
In this figure, the distributions of model parameters are shown as the
two-dimensional intensity color maps.  Any location on the graphical
user interface (\emph{left}) can be clicked on to extract the model
spectrum overlayed on top of the data spectrum (\emph{right}).  The
relatively high-\snrpp\ spectrum yields well-behaving parameter
distributions here.  The figure also gives an example of typical model
parameter distributions.  On the left, the parameters plotted are
$\tau_0$-$\lambda_{\rm red}$ (\emph{top left}), $b_D$-$\lambda_{\rm
red}$ (\emph{top right}), $\Cf$-$\lambda_{\rm red}$ (\emph{bottom
left}), and $\Cf$-$\tau_0$ (\emph{bottom right}).  If marginalized
over the parameter against which they are plotted, the distributions
of $\lambda_{\rm red}$ and $b_D$ become roughly Gaussian, where as
those of $\Cf$ and $\tau_0$ are not.  It is apparent that \Cf\ is not
well constrained for $\Cf \ga 0.45$, and $\tau_0$, while relatively
well constrained, is highly correlated with \Cf.  See \sect{Modeling
NaID Absorption Lines} for detail.
(\emph{A color version of this figure is available in the online journal.})
}
\label{fig.mcmc}
\end{figure}

The measurement pipeline is built on top of PyMC,\footnote{The code
and documentation are available at http://trichech.us/.} which
implements the Metropolis-Hasting algorithm as an MCMC sampler, and
developed in Python.  The distributions of model parameters are
visually inspected with a graphical user interface (GUI) along various
dimensions; see \fig{fig.mcmc}.  The integrity of fitting to the data
spectrum is checked at several interactively picked points on the
two-dimensional intensity map of the distributions of model
parameters.  The GUI-driven visual inspection guards against the fits
latching on to low-\snr\ features and helps to identify unphysical fit
results.  The distribution of $\lambda_{\rm red}$, marginalized over
all the other parameters and convolved with the redshift uncertainty,
needs to be roughly Gaussian and well bounded within $\pm
700~\km~\s^{-1}$ to make it into the high-\snr\ \NaID\ velocity sample
(\sect{Modeling NaID Absorption Lines}).  {\hl The software may be
open sourced at a later date.}

\end{document}